\documentclass[journal]{IEEEtran}
\usepackage[pdftex]{graphicx}
\usepackage{amsmath}
\usepackage{amssymb}
\usepackage{blindtext}
\usepackage{color}
\usepackage{epstopdf}
\usepackage[pdfstartview=FitR]{hyperref}

\ifCLASSINFOpdf
\else
\fi

\begin{document}
%
\title{Berry phase, Berry Connection, and Chern Number for a Continuum Bianisotropic Material from a Classical Electromagnetics Perspective}


\author{\IEEEauthorblockN{S. Ali Hassani Gangaraj~\IEEEmembership{Student member,~IEEE}, M\'ario G. Silveirinha,~\IEEEmembership{Fellow,~IEEE}, and George W. Hanson,~\IEEEmembership{Fellow,~IEEE}}

\thanks{Manuscript received xxx; revised xxx.

G. W. Hanson and S. A. Hassani Gangaraj are with the Department of Electrical Engineering, University of Wisconsin-Milwaukee, Milwaukee, Wisconsin 53211, USA (e-mail: george@uwm.edu, ali.gangaraj@gmail.com.
edu).

M\'ario G. Silveirinha is with the Instituto Superior T\'{e}cnico, University of Lisbon
and Instituto de Telecomunica\c{c}\~{o}es, Torre Norte, Av. Rovisco
Pais 1, Lisbon 1049-001, Portugal (e-mail: mario.silveirinha@co.it.pt).}

}

\markboth{Journal of \LaTeX\ Class Files,~Vol.~xxx, No.~xxx, xxx~2015}%
{Shell \MakeLowercase{\textit{et al.}}: Bare Demo of IEEEtran.cls for Journals}
%




\maketitle

\begin{abstract}
The properties that quantify photonic topological insulators (PTIs), Berry phase, Berry connection, and Chern number, are typically obtained by making analogies between classical Maxwell's equations and the quantum mechanical Schr\"{o}dinger equation, writing both in Hamiltonian form. However, the aforementioned quantities are not necessarily quantum in nature, and for photonic systems they can be explained using only classical concepts. Here we provide a derivation and description of PTI quantities using classical Maxwell's equations, we demonstrate how an electromagnetic mode can acquire Berry phase, and we discuss the ramifications of this effect. We consider several examples, including wave propagation in a biased plasma, and radiation by a rotating isotropic emitter. These concepts are discussed without invoking quantum mechanics, and can be easily understood from an engineering electromagnetics perspective.
\end{abstract}


\maketitle

\IEEEdisplaynontitleabstractindextext

%
\IEEEpeerreviewmaketitle

\section{Introduction}\label{Intro}

\IEEEPARstart{P}{}hotonic topological insulators are emerging as an
important class of material (natural or meta) that allow for the
propagation of unidirectional surface waves, immune to
backscattering, at the interface with another medium
\cite{Joannopoulos}-\cite{Rechtsman1}. There are different types of photonic topological materials, but in
this article we focus on the simplest subclass formed by photonic topological
media with a broken time reversal symmetry, sometimes also
designated as Chern-type insulators (the analogs of quantum Hall insulators). The properties of these
materials are quantified by the Berry phase, Berry connection, and an
invariant known as the Chern number \cite{Haldane}-\cite{Xiao}.
Berry phase was first proposed in 1984 \cite{Berry} for quantum
systems undergoing cyclic evolution. While the occurrence of a phase
factor as a quantum system evolved was long known in quantum
mechanics, it was thought to be non-observable since a
gauge-transformation could remove it. It was Berry who showed that
for a cyclic variation, and assuming adiabatic (sufficiently-slow)
evolution, the phase is not removable under a gauge transformation
\cite{Berry}, and is also observable. It is a form of geometric
phase, related to parallel transport of a vector along a curved
surface \cite{griffithsA}. The concept has been generalized for
non-adiabatic evolution \cite{Aharonov}. In quantum mechanics, a
prototypical example is the progression of electron spin
as a magnetic field vector rotates in a cyclical manner
\cite{griffiths,ballentine}. Here, we are interested in electromagnetic propagation, and the
Berry phase that a classical electromagnetic wave acquires
as it propagates (wave propagation being a form of cyclic evolution,
e.g., the field polarization returning to its initial position after
every wavelength of propagation). See \cite{a}-\cite{e} for some reviews of optical phenomena related to Berry phase. Although not discussed here, we mention another effect related to the Berry phase, the spin Hall effect of light, which is a geometric Berry-phase counterpart of the Lorentz force. This effect has been extensively studied and experimentally verified \cite{y}-\cite{z}.

Often, Berry properties are obtained using the quantum mechanical derivation based on a
Hamiltonian, and making an analogy between Maxwell's equations and
the quantum system. However, this approach has two drawbacks: it
necessitates knowledge of quantum mechanics, and, more importantly,
it obscures the true nature of the phenomena, which in this case is
classical.

There have been previous works considering Berry quantities for
electromagnetics from a classical perspective \cite{b}-\cite{g}, which is intimately connected with
spin-orbit interactions in light \cite{Bliokh}, and from a relativistic wave-equation perspective in \cite{h}-\cite{i}. However, many previous classical works invoke the theory of Hermitian line bundles or gauge
theory (see, e.g., \cite{c}, \cite{Wood}), and here we avoid those topics and consider Berry properties
from a relatively simple electromagnetic perspective \cite{Berry2}.
Furthermore, as described in \cite{Berry3}, before the Berry phase
was understood as the general concept it is now, in various fields
this extra phase had been found. In electromagnetics, the most
notable discoveries were by Pancharatnam in 1956 \cite{Pach}, who
studied polarization evolution of light, and Budden and Smith
\cite{Smith1}-\cite{Smith2}, who considered propagation through the
ionosphere, modeled as an inhomogeneous medium whose parameters
varied gradually with height. Further, pioneering works were done by Rytov in 1938, \cite{j}, and Vladimirskii in 1941, \cite{k}, describing geometrical properties of the polarization evolution of light along curvilinear trajectories; the spin-redirection geometric phase in optics discussed here is sometimes called the Rytov-Vladimirskii-Berry phase. The topic of parallel transport of the polarization of light in an inhomogeneous medium was discussed as early as 1926 by Bortolotti \cite{1926}.

Thus, we can categorize the geometric phases in optics into two classes: (i) The spin-redirection/Rytov--Vladimirskii--Berry) geometric phase, which is the phase associated with changes in wave momentum (with a conserved polarization state), and is the subject of the present work, and (ii) the Pancharatnam--Berry phase,  which appears  when the polarization state varies on the Poincare sphere (while the momentum is unchanged). Here, we consider electromagnetic propagation through a uniform, homogeneous medium, in order to consider perhaps the simplest possible example where important Berry effects occur.

After the discovery of the Berry phase, an early example of a
photonic systems exhibiting such an effect was given in
\cite{Chiao}-\cite{TC1986}, which considered the rotation of the
polarization vector of a linearly polarized laser beam traveling
through a single, helically wound optical fiber. More recently, and
of more direct interest here, there has been a lot of work on
various photonic systems that demonstrate non-trivial Berry
properties, both for periodic and continuum materials
\cite{Solja4}-\cite{Mario2}. The principal interest in these systems
is because at the interface between two regions with different Berry
curvatures, a backscattering-immune (i.e., unidirectional) surface
plasmon-polariton (SPP) can propagate. If the operational frequency
is in a common bandgap of the two bulk materials, the SPP is also
immune to diffraction, and so even arbitrarily-large discontinuities
do not scatter energy.

We also note that one-way SPPs at, e.g., biased plasma interfaces have been observed long before the Berry phase concept \cite{SR}--\cite{AI}, and later, although not within the framework of Berry properties \cite{ZR}. One-way SPPs can also be formed at various other interfaces, such as at the domain walls of Weyl semimetals with broken time-reversal symmetry \cite{ZZ}.

One approach to create PTIs is to use two-dimensional (2D) photonic
crystals with degenerate Dirac cones in their band structure
\cite{Haldane}-\cite{Haldane2}. The degeneracy can be lifted by
breaking time-reversal (TR) symmetry, which opens a band gap and
leads to topologically non-trivial photonic bands. Continuum
materials, either homogenized metamaterials or natural materials,
supporting topologically-protected unidirectional photonic surface
states have also recently been shown
\cite{Mario2}-\cite{Arthur}. Unidirectional surface modes at the
interface between a magnetized plasma or magnetized ferrite and a
metal have been recently studied \cite{Hassani2}-\cite{Hassani1},
where time reversal symmetry is broken by applying a static magnetic
field, opening a bandgap and inducing non-trivial Berry properties.

Unidirectional, scattering-immune surface-wave propagation has great
potential for various waveguiding device applications. The aim of
this paper is to derive and explain all Berry properties from a
classical engineering electromagnetic perspective, without
consideration of the usual quantum mechanics derivation, appealing
to analogies between Schr$\ddot{\mathrm{o}}$dinger's equations and
Maxwell's equations, or invoking gauge theories.

\section{Theory}

\subsection{Maxwell's Equations as a Momentum-Dependent Eigenvalue Problem \label{SecMax}}

In order to establish the necessary concepts, we start by considering a
dispersionless material model, but later extend the results to a lossless
dispersive material model.

Source-free Maxwell's equations are
\begin{align}
& \nabla \times \boldsymbol{\mathrm{E}}\left( \mathbf{r},t\right) =-\frac{%
\partial }{\partial t}\boldsymbol{\mathbf{B}}\left( \mathbf{r},t\right)
\notag  \label{Eq:ME} \\
& \nabla \times \boldsymbol{\mathbf{H}}\left( \mathbf{r},t\right) =\frac{%
\partial }{\partial t}\mathbf{D}\left( \mathbf{r},t\right) ,
\end{align}
and working in the momentum-frequency domain ($\partial /\partial
t\rightarrow -i\omega $ and $\nabla \rightarrow i\mathbf{k}$) and
considering a homogeneous, lossless, bianisotropic material with
frequency-independent dimensionless parameters $\boldsymbol{\varepsilon },~%
\boldsymbol{\mu },~\boldsymbol{\xi },~\boldsymbol{\varsigma }$ representing
permittivity, permeability and magneto-electric coupling tensors,
respectively, the constitutive relations are
\begin{equation}
\left(
\begin{array}{c}
\mathbf{D}\left( \mathbf{k},\omega \right)  \\
\boldsymbol{\mathbf{B}}\left( \mathbf{k},\omega \right)
\end{array}%
\right) =\left(
\begin{array}{cc}
\varepsilon _{0}\boldsymbol{\varepsilon } & \frac{1}{c}\boldsymbol{\xi } \\
\frac{1}{c}\boldsymbol{\varsigma } & \mu _{0}\boldsymbol{\mu }%
\end{array}%
\right) \cdot \left(
\begin{array}{c}
\boldsymbol{\mathbf{E}}\left( \mathbf{k},\omega \right)  \\
\boldsymbol{\mathbf{H}}\left( \mathbf{k},\omega \right)
\end{array}%
\right) .
\end{equation}%
Defining $\boldsymbol{f}_{n}=\left( \boldsymbol{\mathrm{E}}~~\boldsymbol{%
\mathrm{H}}\right) ^{\text{T}}$,
\begin{equation}
\boldsymbol{\mathrm{M}}=\left(
\begin{array}{cc}
\varepsilon _{0}\boldsymbol{\varepsilon } & \frac{1}{c}\boldsymbol{\xi } \\
\frac{1}{c}\boldsymbol{\varsigma } & \mu _{0}\boldsymbol{\mu }%
\end{array}%
\right) ,~\boldsymbol{{\mathrm{N}}}=\left(
\begin{array}{cc}
0 & \boldsymbol{\mathbf{k}}\times \boldsymbol{\mathrm{I}}_{3\times 3} \\
-\boldsymbol{\mathbf{k}}\times \boldsymbol{\mathrm{I}}_{3\times 3} & 0%
\end{array}%
\right) ,   \label{matMN}
\end{equation}%
where $\boldsymbol{{\mathrm{N}}}$ and $\boldsymbol{\mathrm{M}}$ are
Hermitian (the latter since we consider lossless media), we can write
Maxwell's equations as a standard eigenvalue problem,
\begin{equation}
\mathbf{H}\left( \mathbf{k}\right) \cdot \boldsymbol{f}_{n,\mathbf{k}}=\omega _{n,\mathbf{k}}%
\boldsymbol{f}_{n,\mathbf{k}}.  \label{Eq:eigen}
\end{equation}%
where $\mathbf{H}\left( \mathbf{k}\right) =\boldsymbol{\mathbf{M}}^{-1}\cdot
\mathrm{\mathbf{N}}\left( \mathbf{k}\right) $. The electromagnetic
eigenfields are of the form $\tilde{\boldsymbol{f}}_{n}(\mathbf{r})=%
\boldsymbol{f}_{n,\mathbf{k}}e^{i\mathbf{k}\cdot \mathbf{r}}$, where $\boldsymbol{f}_{n,\mathbf{k}}
$ (the solution of Eq. \ref{Eq:eigen}) is the envelope of the fields
(independent of position). In the following we use $\boldsymbol{f}_{n,\mathbf{k}}=\boldsymbol{f}%
_{n}\left( \mathbf{k}\right) $ interchangeably.

The matrix $\mathbf{H(\mathbf{k})}$ is not itself Hermitian even through
both $\boldsymbol{\mathbf{M}}^{-1}$ and $\mathrm{\mathbf{N}}$ are Hermitian,
since $\boldsymbol{\mathbf{M}}^{-1}$ and $\mathrm{\mathbf{N}}$ do not
commute. However, viewed as an operator, $\mathbf{H}$ is Hermitian under the
inner product \cite{Mario1}
\begin{equation}
\left\langle \boldsymbol{f}_{m}|\boldsymbol{f}_{n}\right\rangle =\boldsymbol{%
f}_{n}^{\dagger }(\boldsymbol{\mathbf{k}})\cdot \boldsymbol{\mathrm{M}}\cdot
\boldsymbol{f}_{m}(\boldsymbol{\mathbf{k}}),  \label{Eq:normalization}
\end{equation}%
where the superscript $\dagger $ denotes the conjugate transpose matrix.
Thus, by defining a new set of eigenvectors \cite{Haldane}
\begin{equation}
\mathbf{w}_{n,\mathbf{k}}=\boldsymbol{\mathbf{M}}^{1/2}\boldsymbol{f}_{n,\mathbf{k}}  \label{wf}
\end{equation}%
and the inner product
\begin{equation}
\left\langle \mathbf{w}_{m}|\mathbf{w}_{n}\right\rangle =\mathbf{w}%
_{n}^{\dagger }(\boldsymbol{\mathbf{k}})\cdot \mathbf{w}_{m}(\boldsymbol{%
\mathbf{k}}),  \label{ip1}
\end{equation}%
then%
\begin{equation}
\widetilde{\mathbf{H}}\left( \mathbf{k}\right) \cdot \mathbf{w}%
_{n,\mathbf{k}}=\omega _{n,\mathbf{k}}\mathbf{w}_{n,\mathbf{k}}  \label{eigen2}
\end{equation}%
forms a Hermitian eigenvalue problem, where
\begin{equation}
\widetilde{\mathbf{H}}\left( \mathbf{k}\right) =\boldsymbol{\mathbf{M}}%
^{1/2}\ \boldsymbol{\mathbf{H(\mathbf{k})~M}}^{-1/2} \label{HEP}
\end{equation}%
(that is, $\widetilde{\mathbf{H}}\left( \mathbf{k}\right)$ is an
Hermitian matrix). In the following, because of (\ref{wf}), we can
work with either the eigenfunctions $\mathbf{w}_{n,\mathbf{k}}$ or
$\boldsymbol{f}_{n,\mathbf{k}}$. It is crucially important in what
follows to note that (\ref{eigen2}) and the normalization condition
$\left\langle \mathbf{w}_{n}|\mathbf{w}_{n}\right\rangle=1$ define
the eigenmodes only up to a phase factor.

Maxwell's equations in the form (\ref{eigen2}) is a Hermitian eigenvalue
equation with eigenvalue $\omega _{n,\mathbf{k}}$ and eigenvector $\mathbf{w}_{n,\mathbf{k}}$
which contains the electric and magnetic fields, and, hence, the
polarization. The matrix $\widetilde{\mathbf{H}}\left( \mathbf{k}\right) $
plays the role of the Hamiltonian in quantum mechanics, i.e., it describes
how the systems evolves as momentum changes. In considering the development
of various Berry properties, what is important is to have an eigenproblem
involving two or more related quantities (here we have momentum and
polarization). Although momentum and polarization can be put on an equal
footing \cite{Wood}, it is usually easier to consider momentum as the
parameter, and consider how polarization changes with momentum.

Briefly stated, the Berry phase is the cumulative effect of the relative phase difference between an eigenfunction at $\mathbf{k}$ and  at a nearby point $\mathbf{k}+d\mathbf{k}$. Therefore, if the system changes momentum from some initial value $\mathbf{k}_{i}$ to some final value $%
\mathbf{k}_{f}$, the normalized eigenfunctions change correspondingly; $\mathbf{w}_{n}(%
\mathbf{k}_{i})\rightarrow \mathbf{w}_{n}(\mathbf{k}_{f})$. If the final
momentum equals the initial momentum, then $\widetilde{\mathbf{H}}\left( \mathbf{k}%
_{i}\right) =\widetilde{\mathbf{H}}\left( \mathbf{k}_{f}\right) $
and the system environment returns to its initial value. However,
the eigenfunction may not return to its initial value since
eigenfunctions are defined only up to a phase
factor, and for the case $\mathbf{k}_{i}=\mathbf{k}_{f}$ it may occur that $%
\mathbf{w}_{n}(\mathbf{k}_{f})=e^{i\gamma _{n}}\mathbf{w}_{n}(\mathbf{k}_{i})
$. This anholonomy is represented by a possible additional phase factor, where $\gamma _{n}$ is
called the Berry phase. Berry phase and related quantities in a classical
electromagnetics context is the topic of this work.

\begin{figure}[ht]
    \begin{center}
        \noindent
        \includegraphics[width=3.5in]{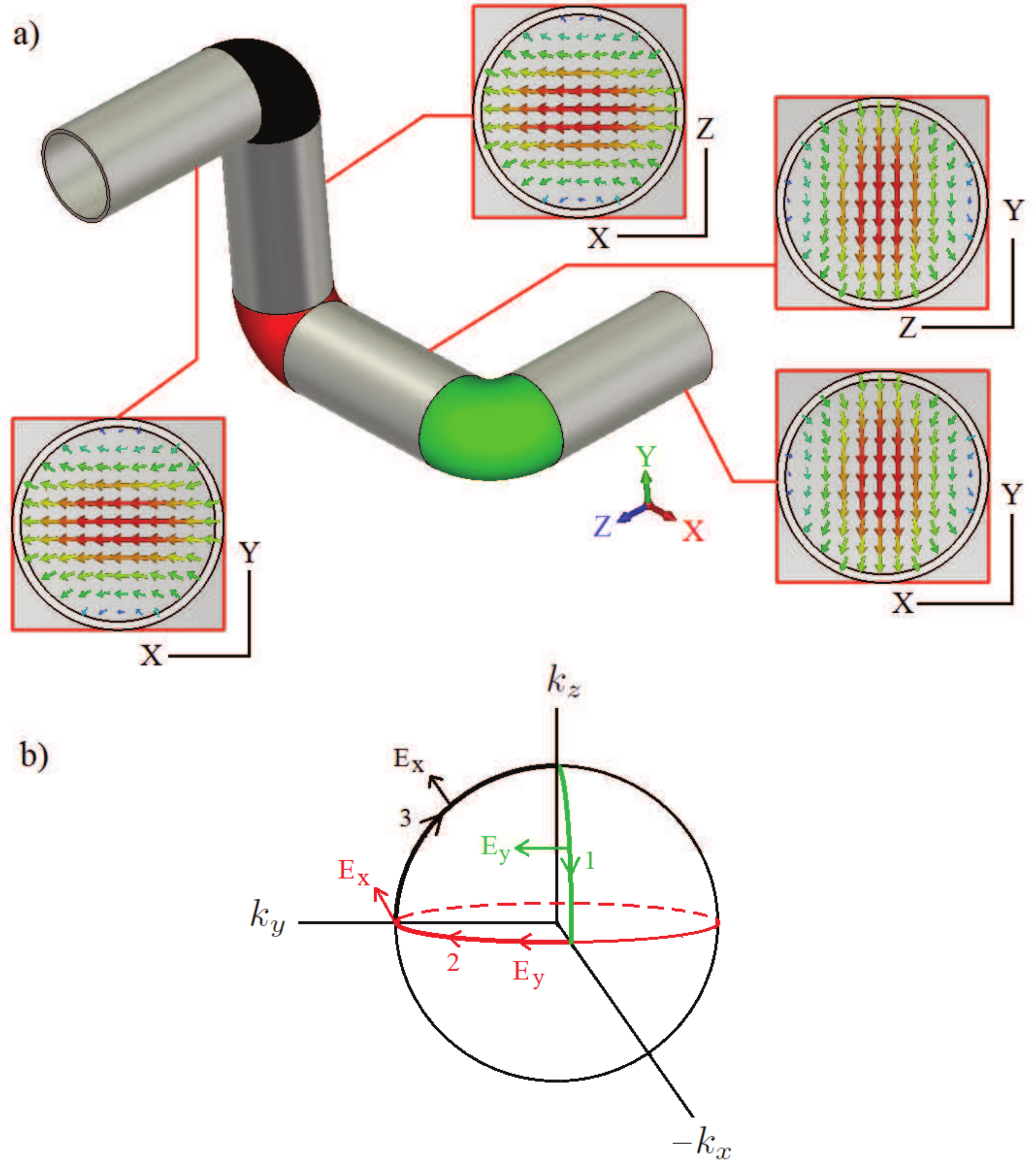}
        \caption{(a) Curved circular waveguide demonstrating polarization rotation due to Berry phase effects, (b) momentum sphere.}\label{Fig1}
    \end{center}
\end{figure}

A first simple example of a non-trivial Berry phase, one in which
the effect is quite evident, arises from considering a curved
circular waveguide supporting the dominant $ \mathrm{TE}_{11} $
mode, as shown in Fig. \ref{Fig1}-a (similar to the case of the
helically-wound optical fiber considered in
\cite{Chiao}-\cite{TC1986}. See also \cite{L}, which considered this effect before the Berry phase was understood.). As the mode propagates it follows the
waveguide, retaining the $ \mathrm{TE}_{11} $ profile. During this
evolution the magnitude of the mode momentum, $ {k_{\mathrm{TE}}}^2
= {k}_x^2 + {k}_y^2 +{k}_z^2 $, is fixed, while its direction
changes, thus the mode traverses a path on the surface of the
momentum sphere depicted in Fig. \ref{Fig1}-b. Note that the
direction of propagation for each point of the path is normal to the
sphere, and thus the electric field is necessarily tangent to the
sphere. At the first bend, the direction of momentum changes from
$z$ toward $x$ (traversing the green path on the momentum sphere),
at the second bend the momentum rotates from $ x $ toward $y$
(traversing the red path on the momentum sphere), and at the last
bend the momentum changes back to its initial direction, $z$
(traversing the black path on the momentum sphere). Thus, the
momentum has traversed a closed path on the momentum sphere. During
this evolution, the polarization (locked to be transverse to the
momentum) changes from $y$ to $x$.

In this case the change in polarization is described by a geometric
phase, the Berry phase, due to parallel-transport of a vector on a
curved surface (a mathematical description is provided later, see also \cite{Berry2}). The phase determines the angle between the initial and final polarizations and depends on the area subtended by the closed path (the phase is geometrical);
Indeed, when a vector is parallel transported along a closed path,
the angle between the initial and final vectors is given by the
integral of the \emph{Gaussian curvature} over the surface enclosed
by the path \cite{Frankel}. For a sphere this corresponds to the
solid angle $\Omega$ subtended by the path.
In this example, $\Omega_\text{sphere}=4 \pi$, and so the subtended
angle is $\Omega_\text{sphere}/8=\pi/2$, which represents the change
from $y$ to $x$ polarization. Traversing a closed path along a non-curved surface does not lead to
this additional angle, and so we see that a non-zero Berry phase has
its origin in the curvature of momentum space  (in the present case
the Gaussian curvature). Furthermore, the geometric-phase evolution of the polarization of light shown in Fig. 1 is valid when the ``adiabatic approximation" holds true, i.e., the helicity (degree of circular polarization) is conserved in the evolution. Non-adiabatic corrections were discussed by Ross \cite{L} and Berry \cite{Berry2}.

Note that, for this case of transverse polarization, the
polarization is normal to the momentum (i.e., we have the
transversality condition $ \boldsymbol{\mathbf{k}} \cdot
\boldsymbol{\hat{\mathrm{p}}} = 0 $, where $
\boldsymbol{\hat{\mathrm{p}}} $ is the polarization unit vector),
and so at any point on the momentum sphere the polarization is
always tangential to the momentum sphere in Fig. \ref{Fig1}-b,
leading to rotation of the polarization as the momentum evolves
along a closed contour. If we were
to consider a waveguide supporting a mode with a longitudinal
component, the polarization of the longitudinal component is aligned
along the direction of momentum (i.e., $ \boldsymbol{\mathbf{k}}
\times \boldsymbol{\hat{\mathrm{p}}} = 0 $), and so is always normal
to the momentum sphere; in this case, as momentum evolves this
component of polarization will not incur any additional phase.

\subsection{Berry Quantities from an Electromagnetic Perspective}

In momentum ($\mathbf{k}$) space, we suppose that the eigenmode $\mathbf{w}%
_{n}(\mathbf{k})$ is initially at some point $\mathbf{k}_{i}$. In the
quantum mechanical case one considers evolution of a system, implicitly as
time progresses, during which some time-dependent parameter comes back to
its initial value. Since here we explicitly consider time-harmonic wave
phenomena, rather then system evolution we simply consider wave propagation
along some path such that the mode ends up at $\mathbf{k}_{f}$. If we
suppose that the traversed path is closed, then $\mathbf{k}_{i}=\mathbf{k}%
_{f}$. Assuming well-defined and single-valued eigenmodes we have the
boundary condition $\mathbf{w}_{n}(\mathbf{k}_{i})=e^{i\gamma _{n}(\mathbf{k})}\mathbf{w}_{n}(\mathbf{k}%
_{f})$. At every point $\mathbf{k}$ on the path, the eigenvalue problem
defines the mode up to a phase factor, which can depend on the momentum and
mode index. We consider two continuous
points $\mathbf{k}$ and $\mathbf{k}+d\mathbf{k}$ and corresponding
eigenmodes $\mathbf{w}_{n}(\mathrm{\mathbf{k}})$ and $\mathbf{w}_{n}(\mathrm{%
\mathbf{k}}+d\mathrm{\mathbf{k}})$. We define the differential phase $%
d\gamma _{n}$ between the two eigenmodes as \cite{Grosso}

\begin{equation}
e^{i(d\gamma _{n})}=\frac{\left\langle \mathbf{w}_{n}(\mathbf{k}+d%
\mathbf{k})|\mathbf{w}_{n}(\mathbf{k})\right\rangle }{\left\vert
\left\langle \mathbf{w}_{n}(\mathbf{k}+d\mathbf{k})|\mathbf{w}%
_{n}(\mathbf{k})\right\rangle \right\vert }=\frac{\mathbf{w}_{n}^{\dagger }(%
\mathbf{k}+d\mathbf{k})\cdot \mathbf{w}_{n}(\mathbf{k})}{\left\vert \mathbf{w%
}_{n}^{\dagger }(\mathbf{k}+d\mathbf{k})\cdot \mathbf{w}_{n}(\mathbf{k}%
)\right\vert }. \label{Eq:aux_Berry_phase}
\end{equation}%
The denominator is present as a normalization factor, and if the modes are
already normalized the denominator is unity since $\left\vert d\boldsymbol{%
\mathrm{k}}\right\vert \ll \left\vert \boldsymbol{\mathrm{k}}\right\vert $.
The above definition is intuitive; if the eigenfunction has a $\mathbf{k}$%
-dependent phase, $\mathbf{w}_{n}(\mathbf{k})=e^{i\zeta \left( \mathbf{k}%
\right) }\boldsymbol{\mathrm{g}}_{n}(\mathbf{k})$, then $\mathbf{w}_{n}(\mathbf{k}+d%
\mathbf{k})=e^{i\zeta \left( \mathbf{k}+d\mathbf{k}\right) }\boldsymbol{\mathrm{g}}_{n}(\mathbf{k}+d\mathbf{k})$, and so
\begin{align}
\mathbf{w}_{n}^{\dagger }(\mathbf{k}+d\mathbf{k})\cdot \mathbf{w}_{n}(%
\mathbf{k})& =e^{-i\zeta \left( \mathbf{k}+d\mathbf{k}\right) }\boldsymbol{\mathrm{g}}_{n}^{\ast }(\mathbf{k}+d\mathbf{k})\cdot e^{i\zeta \left( \mathbf{k}\right)
}\boldsymbol{\mathrm{g}}_{n}(\mathbf{k})  \notag \\
& =e^{-i\zeta \left( d\mathbf{k}\right) }\left( \boldsymbol{\mathrm{g}}_{n}^{\ast }(%
\mathbf{k}+d\mathbf{k})\cdot \boldsymbol{\mathrm{g}}_{n}(\mathbf{k})\right)   \notag
\\
& =e^{-i\zeta \left( d\mathbf{k}\right) }
\end{align}
assuming $\boldsymbol{\mathrm{g}}_{n}$ are normalized, in the limit $\left\vert d%
\mathbf{k}\right\vert \rightarrow 0$.

We expand $\mathbf{w}_{n}(\mathbf{k}+d\mathbf{k})$ in a Taylor series up
to first order in $d\mathbf{k}$ and we expand the exponential to first
order. Considering
\begin{align}
& \nabla _{\boldsymbol{\mathbf{k}}}\left\langle \mathbf{w}_{n}|\mathbf{w}%
_{n}\right\rangle =\nabla _{\boldsymbol{\mathbf{k}}}\mathbf{w}_{n}^{\dagger
}\cdot \mathbf{w}_{n}+\mathbf{w}_{n}^{\dagger }\cdot \nabla _{\boldsymbol{%
\mathbf{k}}}\mathbf{w}_{n}=0  \label{ct} \\
& \rightarrow \nabla _{\boldsymbol{\mathbf{k}}}\mathbf{w}_{n}^{\dagger
}\cdot \mathbf{w}_{n}=-\mathbf{w}_{n}^{\dagger }\cdot \nabla _{\boldsymbol{%
\mathbf{k}}}\mathbf{w}_{n} \notag
\end{align}%
we obtain
\begin{equation}
1+id\gamma _{n}=\mathbf{w}_{n}^{\dagger }(\mathbf{k})\cdot \mathbf{w}_{n}(%
\mathbf{k})-\mathbf{w}_{n}^{\dagger }(\mathbf{k})\cdot \nabla _{\mathbf{k}}%
\mathbf{w}_{n}(\mathbf{k})\cdot d\mathbf{k} \label{TS1}
\end{equation}%
such that $d\gamma _{n}=i\mathbf{w}_{n}^{\dagger }(\mathrm{k})\cdot \nabla _{%
\mathrm{k}}\mathbf{w}_{n}(\mathrm{k})\cdot d\boldsymbol{\mathrm{k}},$ and we
have
\begin{equation}
\gamma _{n}=\oint_{C}d\boldsymbol{\mathrm{k}}\cdot \boldsymbol{\mathrm{A}}%
_{n}(\mathbf{k}).  \label{Eq:Berry_phase}
\end{equation}
The quantity 
\begin{equation}
\boldsymbol{\mathrm{A}}_{n}(\mathbf{k})=i\mathbf{w}%
_{n}^{\dagger }(\mathbf{k})\cdot \nabla _{\mathbf{k}}\mathbf{w}_{n}({\mathbf{%
k}})=i\boldsymbol{f}_{n}^{\dagger }(\mathbf{k})\cdot \mathbf{M}\cdot \nabla
_{\mathbf{k}}\boldsymbol{f}_{n}({\mathbf{k}})
\end{equation}
is called the Berry connection, since it connects the eigenmode
$\mathbf{w}_{n,\mathbf{k}}$ at point $\mathbf{k}$ and at point $\mathbf{k}+d\mathbf{k}$%
; connections arise naturally in gauge theories \cite{Josef}. This is also called the Berry vector potential, as an analogy to the magnetic vector potential. Because
eigenmode $\mathbf{w}_{n}(\mathbf{k})$ is the envelope of the electromagnetic field, in
calculating the Berry phase the usual electromagnetic propagator $e^{i%
\boldsymbol{\mathbf{k}}\cdot \mathbf{r}}$ does not contribute, and
the extra phase $\gamma _{n}$ is a result of the curved geometry of
momentum space.  

Therefore in addition to all other phases, the electric and magnetic fields
can acquire an additional phase $e^{i\gamma _{n}(\mathbf{k})}$ during
propagation, due to anholonomy in momentum space (from (\ref{ct}), it is
easy to show that $\boldsymbol{\mathrm{A}}_{n}(\mathbf{k})$ and $\gamma _{n}$
are real-valued for real-valued $\mathbf{k}$, so that $e^{i\gamma _{n}\left( \mathbf{k}\right) }$ is a
phase, not a decay term). 

\subsubsection{Gauge Considerations and Real-Space Field Analogies}

The electric and magnetic fields (measurable quantities) are invariant under
any electromagnetic gauge transformation. The eigenvalue equation defines
the eigenmodes up to a multiplicative phase factor, which is the well-known
gauge ambiguity of the complex Hermitian eigenproblem. Associated with each
eigenmode is a (gauge) field  in momentum space, which is
the Berry connection.

Multiplication of the eigenfunction $\mathbf{w}_{n}$ by a phase factor (a
unitary transformation) represents a gauge transformation of the Berry
connection; with $\mathbf{w}_{n}\rightarrow e^{\mathrm{i\xi (\mathbf{k})}}%
\mathbf{w}_{n}$, being $e^{\mathrm{i\xi (\mathbf{k})}}$ an arbitrary
smooth unitary transformation,
the Berry connection transforms to
\begin{align}
& \boldsymbol{\mathrm{A}}_{n}^{\prime }=i\mathbf{w}_{n}^{\dagger }(\mathbf{k}%
)e^{-i\mathrm{\xi }(\mathbf{k})}\cdot \nabla _{\mathbf{k}}\left( \mathbf{w}%
_{n}(\mathbf{k})e^{i\mathrm{\xi }(\mathbf{k})}\right)   \notag \\
& ~~~~=i\mathbf{w}_{n}^{\dagger }(\mathbf{k})\cdot \nabla _{\mathbf{k}}%
\mathbf{w}_{n}({\mathbf{k}})-\nabla _{\mathbf{k}}\xi (\mathbf{k})\mathbf{w}%
_{n}^{\dagger }(\mathbf{k})\cdot \mathbf{w}_{n}({\mathbf{k}})  \notag \\
& ~~~~=\boldsymbol{\mathrm{A}}_{n}-\nabla _{\mathbf{k}}\xi (\mathbf{k}),  \label{gauge}
\end{align}%
which means that the Berry connection/vector potential is gauge dependent, like
the electromagnetic vector potential. However, if we consider
paths $C$ that are closed in momentum space, a gauge change in
(\ref{Eq:Berry_phase}) can only change the Berry phase by integer
multiples of $2\pi$ \cite{Hassani1}. Note that the unitary
transformation $e^{\mathrm{i\xi (\mathbf{k})}}$ is required to be a
smooth single-valued function of the wave vector in the vicinity of
the relevant contour $C$, but its logarithm may not be
single-valued, leading to the ambiguity of modulo $2 \pi$ (gauge
dependence) in the Berry phase.

Equation (\ref{Eq:Berry_phase}) is a momentum-space analog to the magnetic flux $%
\Phi _{\text{mag}}$, in terms of the real-space magnetic field and magnetic
vector potential $\boldsymbol{\mathrm{A}}_{\text{mag}}$ in electromagnetics,%
\begin{equation}
\Phi _{\text{mag}}=\int_{S}d\boldsymbol{\mathrm{S}}\cdot \boldsymbol{\mathrm{%
B}}\left( \mathbf{r} \right) =\oint\nolimits_{C}d\boldsymbol{\mathrm{l}}%
\cdot \boldsymbol{\mathrm{A}}_{\text{mag}}\left( \mathbf{r}\right) .
\label{MF1}
\end{equation}%
A corresponding phase for a charged particle in a magnetic vector-potential can be obtained upon multiplying $\Phi _{\text{mag}}$ by $e/\hbar c$, and is known as the Dirac phase. This phase was first described in \cite{O} and underpins the Aharonov--Bohm effect \cite{Berry} (the real-space analog to the Berry phase). 

The real-space magnetic flux density is obtained as the curl of the vector
potential,
\begin{equation}
\boldsymbol{\mathrm{B}}\left( \mathbf{r}\right) =\nabla _{\mathbf{r}}\times
\boldsymbol{\mathrm{A}}_{ \text{mag}}(\mathbf{r}),
\end{equation}%
and, similarly, a momentum-space vector field can be obtained from the curl
of the Berry vector potential/Berry connection,
\begin{equation}
\boldsymbol{\mathrm{F}}_{n}(\mathbf{k})=\nabla _{\mathbf{k}}\times
\boldsymbol{\mathrm{A}}_{n}(\mathbf{k}) \label{Berry_curvature1}
\end{equation}
This field is called the Berry curvature, and can be viewed as an
effective magnetic field in momentum space. In contrast to Berry connection,
this field is clearly gauge-independent. When the Berry
connection is smoothly defined inside the surface enclosed by some
contour $C$ it follows from Stoke's theorem that $\oint\limits_C
{{{\bf{A}}_n}\left( {\bf{k}} \right) \cdot d{\bf{l}}}  =
\int\limits_S {{{\bf{F}}_n}\left( {\bf{k}} \right) \cdot d{\bf{S}}}
$. In such a case, the Berry phase is completely determined by the
Berry curvature. This result is the counterpart of that discussed
earlier, where the geometric Berry phase is determined by the
Gaussian curvature of the momentum space. Yet, it is important to
highlight that in general ${{\bf{A}}_n}\left( {\bf{k}} \right)$ may
not be globally defined in all space. Moreover, in some systems it
turns out that it is impossible to pick a globally defined smooth
gauge of eigenfunctions (even though the whole space can be covered
by different patches of smooth eigenfunctions), and it is this
property that leads to the topological classification of systems as
further detailed ahead.

As a partial summary, the gauge ambiguity of the Hermitian Maxwell
eigenvalue problem allows for several quantities in momentum space;
a geometric gauge-independent (modulo $2 \pi$)
Berry phase, a gauge-dependent Berry vector potential/Berry connection (analogous to the magnetic
vector potential in real space), and a gauge-independent Berry
curvature (analogous to the real-space magnetic flux density). This
correspondence is depicted in Table 1. These quantities arise, in
part, from the material medium, and in part from the type of
electromagnetic mode being considered.

\begin{table}[h!] \centering%
\begin{tabular}{lll}
\hline
& Real-space ($\mathbf{r}$) & Momentum Space ($\mathbf{k}$) \\ \hline\hline 
& Potential $\mathbf{A}_{\text{mag}}\left( \mathbf{r}\right) $ & Connection $\mathbf{A}%
\left( \mathbf{k}\right) $ \\
& Field $\mathbf{B}\left( \mathbf{r}\right) =\nabla \times \mathbf{A}_{\text{%
mag}}\left( \mathbf{r}\right) $ & Curvature $\mathbf{F}\left( \mathbf{k}\right)
=\nabla \times \mathbf{A}\left( \mathbf{k}\right) $ \\
&Flux $\Phi _{\text{mag}}=

\oint\nolimits_{C}d\boldsymbol{\mathrm{l}}%
\cdot \boldsymbol{\mathrm{A}}_{\text{mag}}\left( \mathbf{r}\right)

$ &Phase $\gamma =\oint_{C}d\boldsymbol{\mathrm{k}}\cdot \boldsymbol{\mathrm{A}}%
(\mathbf{k})$ \\ \hline
\end{tabular}%

\caption{Table 1. Analogy between real-space (EM) and momentum space (Berry) quantities. }\label{Tab1}%
\end{table}%

\subsection{Topological classification}

From an applied electromagnetic perspective, one of the most
important consequences of having non-trivial topological properties is the
occurrence of one-way, back-scattering-immune, surface waves at the
interface between two materials with different Berry properties,
specifically two materials that are topologically different.
The topological classification of a system is determined by the
Chern number. The Chern number is an integer
obtained by integrating the Berry curvature over momentum space,
\begin{equation}
\mathcal{C}_{n}=\frac{1}{2\pi }\oint_{S}d\boldsymbol{\mathrm{S}}\cdot
\boldsymbol{\mathrm{F}}_{n}(\mathbf{k}).  \label{Eq:chern}
\end{equation}
In systems for which it is possible to pick a globally defined
smooth gauge of eigenfunctions, and when the momentum space has no
boundary (e.g., in periodic systems), it follows from Stokes
theorem that the Chern number vanishes. Hence, a nonzero zero Chern
number indicates an obstruction to the application of the Stokes
theorem to the entire momentum space. Notably, when momentum space has no boundary the Chern number is quantized such that $
\mathcal{C}_{n}=0, \pm 1, \pm 2, ...$ is an integer. Being an
integer, $ \mathcal{C}_{n}$ must remain invariant under continuous
transformations, and hence $ \mathcal{C}_{n}$ is a topological
invariant and can be used to characterize different topological
phases. It is worth noting that, in general, Berry phase, connection, and curvature are geometric phenomena, while the Chern number is topological. However, the Berry phase can be topological in 2D systems, where it is quantized and takes values of $0$ or $\pi$, and in 1D systems, where it is also known as the Zak phase \cite{Zak}.

The integer nature of $\mathcal{C}_{n}$ arises from the smoothness of the equivalent Hamiltonian and from the fact that the momentum space has no boundary. In the
following, we will primarily be concerned with EM
propagation in the 2D $(k_{x}, k_{y})$ plane, for which the integration in (\ref%
{Eq:chern}) is over the entire 2D wavenumber plane. Even though the
2D $(k_{x}, k_{y})$ plane does not directly fit into the category of
momentum spaces with no boundary, it can be transformed into a
space with no boundary by including the point $\bf{k}=\infty$.
Within this perspective, the momentum space can be mapped to the
Riemann sphere, which has no boundary \cite{Mario2}. However, even
with such a construction the equivalent Hamiltonian is generally
discontinuous at $\bf{k}=\infty$, and, as a consequence, the Chern
number cannot be guaranteed to be an integer. In practice, this
problem can be fixed by introducing a spatial cut-off in the
material response, as discussed in detail in \cite{Mario2}.

In an electromagnetic continuum it is often possible to choose the
eigenfunctions globally defined and smooth in all space with
exception of the point $\bf{k}=\infty$. In this case,
taking $%
C$ to be the perimeter of the infinite 2D $(k_{x}, k_{y})$ plane
and taking into account that the Berry phase for a contour that
encloses a single point (in this case $\bf{k}=\infty$) is always a
multiple of $2\pi$ it follows that
$1=e^{i\oint\nolimits_{C}d\boldsymbol{\mathbf{k}}\cdot
\boldsymbol{\mathrm{A}}_{n}(\mathbf{k})}=e^{i\int_{S}d\boldsymbol{\mathrm{S}}%
\cdot \boldsymbol{\mathrm{F}}_{n}(\mathbf{k})}$
and so $\int_{S}d\boldsymbol{%
\mathrm{S}}\cdot \boldsymbol{\mathrm{F}}_{n}(\mathbf{k})=2\pi n$.

From an electromagnetic perspective, the Chern number can be
interpreted in an intuitive way. From elementary electromagnetics,
Gauss's law relates the total flux over a closed surface $S$ to the
total charge within the surface,
\begin{equation}
\oint_{S}\varepsilon _{0}\boldsymbol{\mathrm{E}}\left( \mathbf{r}\right)
\cdot d\boldsymbol{\mathrm{S}} =Q^{T}=mq,
\end{equation}%
where, assuming identical charged particles, $m$ is the number of
particles and $q$ the charge of each particle (although often
approximated as a continuum, $Q^{T}$ is quantized). To keep things
simple we will assume a monopole charge of strength $mq$ located at
the origin. The electric field is given by Coulomb's law,
\begin{equation}
\boldsymbol{\mathrm{E}}=\left( \frac{mq}{4\pi \varepsilon _{0}}\right) \frac{%
\boldsymbol{\mathrm{r}}}{ r^{3}}.
\end{equation}%
The magnetic form analogous to Gauss's law,
\begin{equation}
\oint_{S}\boldsymbol{\mathrm{B}}\left( \mathbf{r},t\right) \cdot d%
\boldsymbol{\mathrm{S}}=0,  \label{mgl}
\end{equation}%
indicates that there are no magnetic monopoles. However, if magnetic
monopoles existed, the right side of (\ref{mgl}) would be an integer, $mq_%
\textrm{mag}$), and the magnetic flux would be
\begin{equation}
\boldsymbol{\mathrm{B}}=\left( \frac{mq_\textrm{mag}}{4\pi}\right) \frac{%
\boldsymbol{\mathrm{r}}}{ r^{3}}.
\end{equation}

In momentum space, the flux integral over a closed manifold of the Berry
curvature is quantized in units of $2\pi $, indicating the number of Berry
monopoles within the surface. Berry monopoles are the momentum-space analog
of a magnetic monopole, and serve as a source/sink of Berry curvature, just
as the electric charge monopole $mq$ serves as a source/sink of electric
field. If the Chern number is $n$, the net number of Berry monopoles (whose charges do not cancel each other) is $n$.

The Chern number is a bulk property of a material and
is particularly important because, being an integer, it cannot be
changed under continuous deformations of the system. In particular,
the Chern number
$\mathcal{C}_{\text{gap}}=\sum_{n<n_{g}}\mathcal{C}_{n}$ associated
to a subset of bands with $0<\omega < \omega_{\rm{gap}}$, with
$\omega_{\rm{gap}}$ some frequency in a complete bandgap, is
unaffected by a perturbation of the material response (e.g. by a
change of its structural units) unless the gap is closed. This
property has remarkable consequences. Indeed, let us assume that two
materials that share a common bandgap are topologically different,
so that the Chern numbers associated with the bands below the gap
for each material are different. This means that it is impossible to
continuously transform one of the materials into the other without
closing the gap. Then, if one of the materials (let us say in the
region $y<0$) is continuously transformed into the other material
(let us say in region $y> \delta$) then necessarily somewhere in the
transition region ($0<y< \delta$) the bandgap must close. Thus, the
transition region enables wave propagation, and because the region
$y<0$ and $y>\delta$ share a common band gap, a waveguide is formed,
leading to the emergence of topologically protected edge states. It
turns out that the argument is valid even in the limit $\delta \to
0$ and that the number of edge states traveling in the $+x$ and
$-x$ directions are generally different. Remarkably, the difference
between the Chern numbers of the two materials
($\mathcal{C}_{\text{gap},\Delta
}=\mathcal{C}_{2}-\mathcal{C}_{1}\neq 0$) gives the difference
between the number of states propagating in the two directions. This
property is known as the \emph{bulk-edge correspondence principle}.
Importantly, due to the difference in the number of propagating
states, it is possible to have unidirectional propagation with no
backscattering in a number of physical channels that equals the
Chern number difference. In the case of an electromagnetic continuum
the application of the bulk-edge correspondence requires special
care, namely it is crucial to mimic the high-frequency spatial
cut-off that guarantees that the Chern numbers are integer
\cite{bulkedge}.


\subsection{Isofrequency surfaces in momentum space}\label{s1}

Let us now focus in the interesting case wherein the momentum space
is a isofrequency surface, so that the relevant eigenstates are
associated with the same oscillation frequency $\omega$.

In line with a previous discussion, ${{\bf{w}}_n}\left( {{\bf{k}} +
d{\bf{k}}} \right)$ can always be expanded in terms of the basis
${\left\{ {{{\bf{w}}_m}\left( {\bf{k}} \right)} \right\}_{m =
1,...,n,...}}$. In this manner one obtains an expansion of the form
${{\bf{w}}_n}\left( {{\bf{k}} + d{\bf{k}}} \right) = ... +
{e^{-i{d\gamma _n}}}{{\bf{w}}_n}\left( {\bf{k}} \right) + ....$
(only the leading term of the expansion is shown), being ${d\gamma
_n} = {{\bf{A}}_{n{\bf{k}}}} \cdot d{\bf{k}}$ the Berry phase for
the path determined by $d\bf{k}$. Hence, the projection of
${{\bf{w}}_n}\left( {{\bf{k}} + d{\bf{k}}} \right)$ onto
${{\bf{w}}_n}\left( {{\bf{k}}} \right)$ is determined by the
coefficient ${e^{-i{d\gamma _n}}}$, consistent with
\eqref{Eq:aux_Berry_phase}.

Because it is assumed that the eigenstates oscillate with the same
frequency in real-space, it follows that the infinitesimal Berry
phase $d\gamma _n$ may be regarded as a tiny time advance (or time
delay). In other words, it determines if neighboring eigenstates (in
the same isofrequency surface) oscillate in phase or not. When
$d\gamma _n$ is nonzero the relative difference between the two
neighboring eigenstates in the time domain is minimized when the
eigenstates are calculated at \emph{different} time instants, with a
time advance/delay determined $d\gamma _n$. Within this perspective,
the Berry curvature flux (${\nabla _{\bf{k}}} \times
{{\bf{A}}_{n{\bf{k}}}} \cdot {\bf{\hat n}}ds$) gives the accumulated
Berry phase (accumulated delay, from point to point) for a small
loop in the isofrequency surface with normal $\hat n$.

In the particular case wherein one restricts the analysis to the
2D-plane, the isofrequency surface reduces to some closed contour.
The corresponding Berry phase is determined (modulo $2 \pi$) by the
integral of the Berry curvature over the surface enclosed by the
loop, and determines the accumulated phase delay between neighboring
eigenstates.

We can now revisit the first example of a curved waveguide, Fig. \ref{Fig1}, from a
mathematical perspective. Using $\mathbf{A}_{n\mathbf{k}}=i%
\boldsymbol{f}_{n\mathbf{k}}^{\dagger }\cdot \mathbf{M}\cdot \nabla _{%
\mathbf{k}}\boldsymbol{f}_{n\mathbf{k}}$ where $\mathbf{M}=\mathrm{diag}%
\left( \varepsilon _{0\text{ 3x3}},\mu _{0\text{ 3x3}}\right) $ and assuming a circularly polarized (CP) field envelope 
\begin{equation}
\mathbf{f}_{n\mathbf{k}}^{\pm }=\frac{1}{\sqrt{2\varepsilon _{0}}}\left( 
\begin{array}{c}
\mathbf{e}_{\pm } \\ 
\mp \frac{1}{\eta _{0}}i\mathbf{e}_{\pm }%
\end{array}%
\right) \label{cp1}
\end{equation}%
(where $\eta _{0}=\sqrt{\mu _{0}/\varepsilon _{0}}$), with $\mathbf{e}%
_{\pm }=\left( \widehat{\mathbf{\theta }}\pm i\widehat{\mathbf{\varphi }}%
\right) /\sqrt{2}$, then $A_{\mathbf{k}\theta }^{\pm }=i\boldsymbol{f}_{n\mathbf{k}%
}^{\dagger \pm }\cdot \mathbf{M}\cdot \left( 1/k\right) \partial _{\theta }%
\boldsymbol{f}_{n\mathbf{k}}^{\pm }=0$ and 
\begin{equation}
A_{\mathbf{k}\varphi }^{\pm }=i\boldsymbol{f}_{n\mathbf{k}}^{\dagger \pm }\cdot 
\mathbf{M}\cdot \left( 1/k\sin \theta \right) \partial _{\varphi }\boldsymbol{f}%
_{n\mathbf{k}}^{\pm }=\pm \frac{1}{k\sin \theta }\cos \theta ,
\end{equation}%
leading to 
\begin{equation}
\mathbf{F}_{\mathbf{k}}^{\pm }=\mp \frac{\widehat{\mathbf{k}}}{k^{2}}. \label{cp3}
\end{equation}%
Thus, we have a Berry monopole at $k=0$, and the radial component of the Berry curvature is precisely the Gaussian curvature of the isofrequency surface. Because the
accumulated phase for each CP state is symmetric (one suffers a time delay,
the other, a time advance), the linearly polarized state in Fig. \ref{Fig1},
decomposed into two counter-propagating CP states, incurs a rotation after
transversing the curved waveguide. Note that in this example the relevant eigenspace is doubly degenerate, which is why for linear polarization the final state can be different from the initial state. For waveguides with a single mode (such as a $\mathrm{TE_{10}}$ mode in a rectangular waveguide, the final state is always identical to the initial state. Finally, we can note that curvature (\ref{cp3}) together with the spherical isofrequency surface in k-space naturally yield nonzero Chern numbers for free-space light ((\ref{cp3}) substituted into (\ref{Eq:chern}) leads to $C_n=\pm2$ for the two helicity states $\pm$ \cite{q}).

Another intuitive physical example is the following. Consider an
hypothetical isotropic emitter radiating with
oscillation frequency $\omega$ in an unbounded electromagnetic
material. The far-field pattern is determined by the isofrequency contour of
the material. For simplicity let us suppose that the isofrequency
contours are spherical surfaces. In that case, the
far-field in the direction $\hat {\bf{r}}$ it determined by a mode
with wave vector parallel to $\hat {\bf{r}}$. Hence, in the
far-field one has ${\bf{f}} =
{{\bf{f}}_{\bf{k}}}{e^{i{\bf{k}} \cdot {\bf{r}}}}/r$, where
$\bf{k}$ depends on the observation direction, and
${{\bf{f}}_{\bf{k}}}$ is the envelope of a plane wave propagating
along $\bf{k}$. Because the radiator is isotropic, the modes
${{\bf{f}}_{\bf{k}}}$ have all the same normalization (the
3D radiation pattern intensity is independent of
the observation direction). Hence, for example, the
$\varphi$ component of the Berry connection (tangent to the
isofrequency contour) determines the differential phase delay
between the emitted fields for neighboring observation directions
$\varphi$ and $\varphi + d\varphi$. Note that the
Berry connection may depend on the considered isotropic emitter. Moreover, the relevant gauge is fixed by the emitter.

To further illustrate the idea and discuss the implications of a
nontrivial Berry phase, let us consider the case where the relevant
medium is air and that the isotropic emitter has a far-field determined
by $\bf{f}_{\bf{k}} \equiv {{\bf{f}}_\mathit{E}}\left( {{\bf{\hat r}}}
\right)$. Note that in the present problem ${{\bf{\hat k}}}$ can be identified
with ${{\bf{\hat r}}}$. Because the emitter is isotropic one has $\left\langle {{{\bf{f}}_{\bf{k}}}|{{\bf{f}}_{\bf{k}}}}
\right\rangle = const.$, independent of the observation direction.

Let us suppose that a \emph{polarization matched} receiving antenna is placed in the far-field of
the emitter, let us say along the direction $\theta  = \pi /2, \, \varphi  = 0$. Then, the voltage induced at the open terminals of the
receiver is of the form:
\begin{equation}
\begin{array}{l}
{V_{oc}} \equiv {V_{oc,0}} =  - {\bf{h}}_e^R \cdot {{\bf{E}}_E}\\
{\rm{     }} =  - {\left. {\left( {h_{e,\theta }^R{E_{E,\theta }} + h_{e,\varphi }^R{E_{E,\varphi }}} \right)} \right|_{\varphi  = 0,\theta  = \pi /2}}
\end{array}
\end{equation}
where ${\bf{h}}_e^R$ is the (vector) effective length of the receiving antenna and ${{\bf{E}}_E}$ is the incident electric field. Since the receiving antenna is polarization matched there is some constant $A$ such that ${\bf{h}}_e^R = A{\bf{E}}_E^*$.

Consider that the isotropic emitter can rotate around the polar ($z$) axis. When the emitter is rotated by an angle $d\varphi$ the induced voltage at the terminals of the emitter is ${V_{oc}} =  - \left( {{{\left. {h_{e,\theta }^R} \right|}_{\varphi  = 0}}{{\left. {{E_{E,\theta }}} \right|}_{\varphi  =  - d\varphi }} + {{\left. {h_{e,\varphi }^R} \right|}_{\varphi  = 0}}{{\left. {{E_{E,\varphi }}} \right|}_{\varphi  =  - d\varphi }}} \right)$, so that for a small angle it is possible to write ${V_{oc}} = {V_{oc,0}} + \delta {V_{oc}}$ with
\begin{equation}
\delta {V_{oc}} = {\left. {\left( {h_{e,\theta }^R{\partial _\varphi }{E_{E,\theta }} + h_{e,\varphi }^R{\partial _\varphi }{E_{E,\varphi }}} \right)} \right|_{\varphi  = 0,\theta  = \pi /2}}d\varphi. 
\end{equation}
Taking into account that for $\theta  = \pi /2, \, \varphi  = 0$ one has ${\partial _\varphi }{\bf{\hat \theta }} = 0$ and ${\partial _\varphi }{\bf{\hat \varphi }} =  - {\bf{\hat r}}$ one may write the induced voltage perturbation as (note that ${\bf{h}}_e^R \cdot {\bf{\hat r}} = 0$):
\begin{equation}
\delta {V_{oc}} = {\bf{h}}_e^R \cdot {\partial _\varphi }{{\bf{E}}_E}d\varphi.
\end{equation}
Using the fact that the receiving antenna is polarization matched (when $d\varphi=0$) one gets:
\begin{equation}
{V_{oc}} = {V_{oc,0}}\left( {1 - \frac{{{\bf{E}}_E^* \cdot {\partial _\varphi }{{\bf{E}}_E}}}{{{\bf{E}}_E^* \cdot {{\bf{E}}_E}}}} {d\varphi } \right).
\end{equation}
In free-space the term in brackets can be written as a function of the six-vector ${{\bf{f}}_E}$ that determines the far-field of the isotropic emitter: $\frac{{{\bf{E}}_E^* \cdot {\partial _\varphi }{{\bf{E}}_E}}}{{{\bf{E}}_E^* \cdot {{\bf{E}}_E}}} = \frac{{\left\langle {{{\bf{f}}_E}|{\partial _\varphi }{{\bf{f}}_E}} \right\rangle }}{{\left\langle {{{\bf{f}}_E}|{{\bf{f}}_E}} \right\rangle }}$. Hence, the induced voltage is expressed in terms of the Berry connection associated with ${{\bf{f}}_E}$,
\begin{equation}
{V_{oc}} = {V_{oc,0}}\left( {1 + i{A_\varphi }{k_0}d\varphi } \right).
\end{equation}
Here $k_0 = \omega/c$ is the free-space wave number and $ A_\varphi$ is the azimuthal component of the Berry connection. It is interesting to note that $d\gamma  =  - {A_\varphi }{k_0}d\varphi$ corresponds exactly to the infinitesimal Berry phase (note that as the emitter is rotated by an angle $d\varphi$ the wave vector - from the point of view of the receiver - is effectively rotated by an angle $-d\varphi$). From here one can write ${V_{oc}} \approx {V_{oc,0}}{e^{ - id\gamma }}$, so that in the time domain (without loss of generality, $V_{oc,0}$ is assumed real-valued):
\begin{equation}
{V_{oc}}\left( t \right) \approx {V_{oc,0}}\cos \left( {\omega t + d\gamma } \right)
\end{equation}
The physical interpretation is the following: as the isotropic emitter is rotated around the $z$ axis the polarization matched receiver probes the far-field in a different azimuthal direction. As neighboring states in a isofrequency contour differ approximately by a time-advance determined by the Berry phase the induced voltage gains the phase advance $d\gamma$.

Notably, this discussion shows that a nontrivial Berry phase can have remarkable physical consequences in the considered radiation scenario. Indeed, suppose that the emitter vibrates with frequency $\Omega$ so that $d\varphi$ changes with time as $d\varphi  = d{\varphi _0}\cos \left( {\Omega t} \right)$. In such a case the frequency spectrum of the voltage at the terminals of the matched receiver gains two spectral components due to the voltage perturbation $\delta {V_{oc}}\left( t \right) \approx  - {V_{oc,0}}\sin \left( {\omega t} \right)d\gamma$. Thus, a nontrivial Berry phase leads to the appearance of two spectral lines at $\omega  \pm \Omega$, reveling the analogy between the Berry phase and angular Doppler (or Coriolis) effect \cite{r}-\cite{x}.
This effect can occur only when either ${E_{E,\theta}}$ and/or ${E_{E,\varphi}}$ vary with $\varphi$ in the equatorial line ($\theta=\pi/2$), for example, when the spherical field components have a $\varphi$ dependence of the type ${e^{in\varphi }}$ with $n \ne 0$. Even though in the previous discussion the emitter was assumed to be isotropic, in practice it is enough that the radiation intensity is constant in some solid angle that contains the observation direction ($\varphi=0$ and $\theta = \pi/2$).

Considering, e.g., $\mathbf{E}_E=E_{\theta }\left( \theta ,\varphi \right) \widehat{%
\mathbf{\theta }}$ where $E_{\theta }\left( \theta ,\varphi \right)=h\left(
\theta \right) e^{ig\left( \varphi \right) }$, then $A_{\varphi }=\left(
-1/k_{0}\right) \partial g/\partial \varphi $. Thus, in this case the Berry
phase is proportional to the slope (Berry-phase gradient) \cite{x} of the emitter antenna far-field phase.\\

\subsection{Dispersive Material Model}

\label{1.3}

For the dispersion-less material model Maxwell's equations can be written in
the form of a standard Hermitian eigenvalue problem (\ref{eigen2}). However,
for a dispersive lossless material model this is not the case since $\widetilde{\mathbf{H}}=\widetilde{\mathbf{H}}\left( \mathbf{k},\omega\right)$. This non-standard eigenvalue problem was
considered in \cite{Haldane} for periodic materials (without magneto-optic
coupling parameters), and in \cite{Mario2} for continuum models of dispersive
lossless materials and a subclass of nonlocal materials $\boldsymbol{\mathrm{M}}=\boldsymbol{\mathrm{M}}\left( \omega ,\mathbf{k}%
\left( \omega \right) \right) $
including magneto-optic coupling. The details are fairly involved \cite{Mario2}, yet the result is that
defining the eigenvectors as
\begin{equation}
\mathbf{w}_{n}(\boldsymbol{\mathbf{k}})=\left[ \partial_{\omega} \left( \omega
\boldsymbol{\mathrm{M}}\left( \omega ,\mathbf{k}\right) \right) \right]
^{1/2}\cdot \boldsymbol{f}_{n}(\boldsymbol{\mathbf{k}}) \label{KD}
\end{equation}%
and the inner product $\left\langle \mathbf{w}_{n}|\mathbf{w}%
_{m}\right\rangle =\mathbf{w}_{n}^{\dagger }\cdot \mathbf{w}_{m}=\boldsymbol{%
f}_{n}^{\dagger }(\boldsymbol{\mathbf{k}})\cdot \partial \omega \left(
\omega \boldsymbol{\mathrm{M}}\left( \omega ,\mathbf{k}\right) \right) \cdot
\boldsymbol{f}_{m}(\boldsymbol{\mathbf{k}})$ (both of which reduce to the
dispersionless case, (\ref{wf}) and (\ref{Eq:normalization}) for $%
\boldsymbol{\mathrm{M}}$ a constant matrix) it is possible to define
a standard Hermitian eigenvalue equation in terms of auxiliary
parameters. The end result is that the formulas developed for the
dispersionless case work
for the dispersive case if we replace $\boldsymbol{\mathrm{M}}$ with $%
\partial \omega (\omega \boldsymbol{\mathrm{M}}{(\omega ,\mathbf{k})})$ in (%
\ref{Eq:normalization})-(\ref{HEP}), leading to 
\begin{equation}
\boldsymbol{\mathrm{A}}_{n}(\mathbf{k})=\mathrm{Re}\left\{ i\boldsymbol{f}%
_{n}^{\dagger }(\mathbf{k})\cdot \partial \omega (\omega \boldsymbol{\mathrm{%
M}}{(\omega ,\mathbf{k})})\cdot \nabla _{\mathbf{k}}\boldsymbol{f}_{n}({%
\mathbf{k}})\right\}   
\end{equation}%
where the operator $\mathrm{Re}\left\{ \cdot \right\} $ is unnecessary for the
dispersionless case.

\subsubsection{Symmetry Conditions Leading to Non-Trivial Berry Connection and Curvature}

In the appendix we show that for the lossless case, a reciprocal
medium is a medium with time-reversal symmetry and vice-versa. We
also show that for TR ($\mathcal{T}$) invariant materials,
$\boldsymbol{\mathrm{F}}_{n}(\mathbf{k})=-\boldsymbol{\mathrm{F}}_{n}(\mathbf{-k})$,
and for inversion ($\mathcal{I}$) symmetric materials,
$\boldsymbol{\mathrm{F}}_{n}(\mathbf{k})=\boldsymbol{\mathrm{F}}_{n}(\mathbf{-k})$.

In Eq. \ref{Eq:chern} the integral is taken over the entire plane of
propagation from $\mathbf{k}\rightarrow -\infty $ to $\mathbf{k}\rightarrow
+\infty $ and therefore Berry curvature $\boldsymbol{\mathrm{F}}_{n}$ is
comprised of both forward $(\mathbf{k}>0)$ and backward (time-reversed) $(%
\mathbf{k}<0)$ modes. Given the odd symmetry of
$\boldsymbol{\mathrm{F}}_{n}$ for a $\mathcal{T}$-invariant material, in 2D (e.g., the $(k_{x}, k_{y})$ plane) the $z$ component of the Berry curvature is an odd function of kx and ky, leading to a zero Chern number upon summing over all k values. In this case, in order to have non-zero Chern number one must break time reversal
symmetry using, for example, a non-reciprocal material. However, as detailed in Section \ref{s1}, integration over an equi-frequency surface for free-space light leads to a non-zero Chern number despite the presence of $\mathcal{T}$-invariance. Obviously, if we have both $\mathcal{T}$- and $\mathcal{I}$-symmetry for a considered mode, then $\boldsymbol{\mathrm{F}}_{n}(\boldsymbol{\mathrm{k}})=\mathbf{0}$. 

\section{Example and discussion: Magneto-Plasma materials}

A prominent example of a non-reciprocal medium is obtained when we apply a
magnetic field to a medium with free charges. Consider a magneto-optic
medium characterized by
\begin{equation}
\boldsymbol{\varepsilon }=\varepsilon _{0}\left[
\begin{array}{ccc}
\varepsilon _{11} & i\varepsilon _{12} & 0 \\
-i\varepsilon _{12} & \varepsilon _{11} & 0 \\
0 & 0 & \varepsilon _{33}%
\end{array}%
\right] ,~~\boldsymbol{\mu }=\mu _{0}\boldsymbol{\mathrm{I}}_{3\times 3},
\label{Eq:tensor}
\end{equation}%
which can be realized using a plasma biased with a static magnetic field
along the $z$ direction. The elements of the permittivity tensor are
dispersive,
\begin{equation}
\varepsilon _{11}=1-\frac{\omega _{p}^{2}}{\omega ^{2}-\omega _{c}^{2}}%
,~~\varepsilon _{12}=\frac{-\omega _{c}\omega _{p}^{2}}{\omega \left( \omega
^{2}-\omega _{c}^{2}\right) },~~\varepsilon _{33}=1-\frac{\omega _{p}^{2}}{%
\omega ^{2}}  \label{biased_plasma}
\end{equation}%
where the cyclotron frequency is $\omega _{c}=\left(
q_{e}/m_{e}\right) \mathrm{B}_{z}\ $and the plasma frequency is
$\omega _{p}^{2}=N_{e}q_{e}^{2}/\varepsilon _{0}m_{e}$. In the
above, $N_{e}$ is the free electron density, and $q_{e} = -1.60e-19 $ C and $m_{e} = 9.10 e-31$ kg are the electron charge and mass, respectively. A biased plasma has a
Hermitian permittivity matrix but does not satisfy the requirements
provided in the appendix for being TR invariant.

For a biased magnetoplasma there are two principle configurations for wave
propagation, propagation along the biasing field and propagation
perpendicular to the biasing field. Considering wave propagation in the bulk
of a magnetoplasma medium, the associated electromagnetic waves envelopes
can be obtained by finding the solution, $\boldsymbol{f}_{n}$, of (\ref%
{Eq:eigen}), $\boldsymbol{\mathbf{H(\mathbf{k}}},\omega \boldsymbol{\mathbf{)%
}}\cdot \boldsymbol{f}_{n}=\omega _{n}\boldsymbol{f}_{n}$;
\begin{equation*}
\left(
\begin{array}{cc}
\label{vee}-\boldsymbol{\mathrm{I}}_{3\times 3} & -\frac{\boldsymbol{%
\varepsilon }^{-1}}{\omega _{n}\varepsilon _{0}}\cdot \boldsymbol{\mathrm{k}}%
\times \boldsymbol{\mathrm{I}}_{3\times 3} \\
\frac{1}{\omega _{n}\mu _{0}}\cdot \boldsymbol{\mathrm{k}}\times \boldsymbol{%
\mathrm{I}}_{3\times 3} & -\boldsymbol{\mathrm{I}}_{3\times 3}%
\end{array}%
\right) \cdot \left(
\begin{array}{c}
\boldsymbol{\mathrm{E}} \\
\boldsymbol{\mathrm{H}}%
\end{array}%
\right) =0.
\end{equation*}

\subsection{Propagation Parallel to the Static Bias}

First we suppose that the field propagates along the bias field ($z$%
-direction), which leads to the well-known Faraday rotation. From Faraday's and Ampere's law, assuming a sourceless medium, the electric field is $\boldsymbol{\mathrm{e}}^{\pm }=\mathrm{E}_{x}^{\pm }\left( \hat{%
x}\pm i\hat{y}\right) $, such that CP eigenfunctions are  
\begin{equation}
\boldsymbol{f}^{\pm}_{nk} = \left(%
\begin{array}{c}
\boldsymbol{\mathrm{e}}^{\pm} \\ 
\mp i \mathrm{Y}^{\pm} \boldsymbol{\mathrm{e}}^{\pm}%
\end{array}%
\right)
\end{equation}
where $\mathrm{k}_{\pm }=\mathrm{k}_{0}\sqrt{\varepsilon _{11}\mp \varepsilon _{12}%
}$, $\mathrm{k}_{z}=\left( \mathrm{k}_{+}+\mathrm{k}_{-}\right) /2$, and $%
\mathrm{Y}^{\pm }=\sqrt{\varepsilon _{11}\mp \varepsilon _{12}}/\eta _{0}$ \cite{Ishimaru}. Based on (\ref{cp1})-(\ref{cp3}), it can be seen that for a CP wave traveling along a straight path where $\mathbf{k}$ is constant, there is no Berry phase effect.

\subsection{Propagation Perpendicular to the Static Bias}

For the case of wave propagation perpendicular to the biasing field (no
variation along $z$-direction, $(\partial / \partial z = 0)$), it can be
shown that the modes decouple into a TE mode $(\mathrm{E}_z,~ \mathrm{H}_x,~
\mathrm{H}_y) $ and a TM-mode $(\mathrm{E}_x, ~\mathrm{E}_y,~ \mathrm{H}_z) $%
.

For TE modes, $~~\mathbf{E}=\widehat{\mathbf{z}}E_z \rightarrow \mathbf{H}=%
\frac{\mathbf{k}}{\omega \mu _{0}}\times \widehat{\mathbf{z}}$ such that the
eigenmodes ($6\times 1$ vectors) are
\begin{equation}
\boldsymbol{f}_{nk}^{\text{TE}}=\left(
\begin{array}{c}
\widehat{\mathbf{z}} \\
\frac{\mathbf{k}}{\mu _{0}\omega _{nk}}\times \widehat{\mathbf{z}}%
\end{array}%
\right) \mathrm{E}_z
\end{equation}
and the dispersion equation is $\mathrm{k}_x^2 + \mathrm{k}_y^2 =
\varepsilon _{33} \left(\omega_n/c\right)^2 $.

For TM modes $\boldsymbol{\mathrm{H}}=\hat{\mathbf{z}}\mathrm{H}_z$, $%
\boldsymbol{\mathrm{E}}=\boldsymbol{\varepsilon }^{-1}\cdot \left( \hat{\mathbf{%
z}}\times \boldsymbol{\mathrm{k}}\right) \boldsymbol{/}\left( \omega
_{n}\varepsilon _{0}\right) $,
\begin{equation}
~~\boldsymbol{f}_{nk}^{\text{TM}}=\left(
\begin{array}{c}
\boldsymbol{\varepsilon }^{-1}\cdot \boldsymbol{\hat{\mathbf{z}}}\times \frac{%
\boldsymbol{\mathrm{k}}}{\varepsilon _{0}\omega _{n}} \\
\boldsymbol{\hat{\mathbf{z}}}%
\end{array}%
\right) \mathrm{H}_z
\end{equation}
with dispersion relation, 
\begin{equation}
\mathrm{k}_{x}^{2}+\mathrm{k}_{y}^{2}=\varepsilon _{\text{eff}}\left( \frac{%
\omega _{n}}{c}\right) ^{2},  \label{Eq:TM_bulk}
\end{equation}%
where $\varepsilon _{\text{eff}}=\left( \varepsilon _{11}^{2}-\varepsilon 
_{12}^{2}\right) /\varepsilon _{11}$.

For TE modes there is no Berry phase (in the considered gauge the eigenfunctions are real-valued, and so $\mathbf{A}=\mathbf{F}=\mathbf{0}$). For TM modes, eigenfunctions are complex-valued due to the material permittivity, allowing nontrivial Berry properties. Denoting the frequency derivative of the material response matrix, $%
\partial_{\omega }(\omega \boldsymbol{M})$ as $\beta _{11}=\partial _{\omega
}(\omega\varepsilon _{0}\varepsilon _{11})$, $\beta _{12}=\partial _{\omega
}(\omega \varepsilon _{0} i\varepsilon _{12})$, it can be shown \cite{Hassani1} that
the Berry connection is
\begin{equation}
\boldsymbol{\mathrm{A}}_{n}=\frac{\mathrm{Re}\{i \boldsymbol{f}_{nk}^{\dagger
}\cdot \frac{1}{2} \frac{\partial }{\partial \omega }(\omega \boldsymbol{%
\mathrm{M}}{(\omega )})\partial _{k}\boldsymbol{f}_{nk}\}}{\boldsymbol{f}%
_{nk}^{\dagger }\cdot \frac{1}{2} \frac{\partial }{\partial \omega }(\omega
\boldsymbol{\mathrm{M}}{(\omega )})\boldsymbol{f}_{nk}}=\frac{ \mathrm{Re}\{%
\mathrm{N}_{x} \hat{\mathbf{x}} +\mathrm{N}_{y} \hat{\mathbf{y}} \}}{\mathrm{%
D}}  \label{AB1}
\end{equation}%
where%
\begin{align}
&\mathrm{N}_{x}=\frac{i}{(\varepsilon _{0}\omega _{n})^{2}}\{-2\alpha
_{11}\alpha_{12}[k_{x}\beta _{12}+k_{y}\beta _{11}]  \notag \\
& ~~~~~~ +(|\alpha _{11}|^{2}+|\alpha_{12}|^{2})[k_{x}\beta _{11}-k_{y}\beta
_{12}]\}  \notag  \label{AB2} \\
& \mathrm{N}_{y}=\frac{i}{(\varepsilon _{0}\omega _{n})^{2}}\{2\alpha
_{11}\alpha _{12}[k_{x}\beta _{11}-k_{y}\beta _{12}]  \notag \\
& ~~~~~~ +(|\alpha _{11}|^{2}+|\alpha _{12}|^{2})[k_{x}\beta
_{12}+k_{y}\beta _{11}]\}  \notag \\
& \mathrm{D}=\frac{|k|^{2}}{(\varepsilon _{0}\omega _{n})^{2}}%
[(|\alpha_{11}|^{2}+|\alpha _{12}|^{2})\beta _{11}-2\alpha _{11}\alpha
_{12}\beta_{12}]+\mu_0
\end{align}%
and
\begin{equation}
\alpha _{11}=\frac{\varepsilon _{11}}{\varepsilon _{11}^{2}-\varepsilon _{12}^{2}}%
,~\alpha _{12}=\frac{-i\varepsilon _{12}}{\varepsilon _{11}^{2}-\varepsilon _{12}^{2}}
\end{equation}
and for the Berry curvature,
\begin{equation}
\boldsymbol{\mathrm{F}}_{n}=\frac{\hat{\mathbf{z}}}{\mathrm{D}(\varepsilon 
_{0}\omega _{n})^{2}}\mathrm{Re}\{i\{4\alpha _{11}\alpha _{12}\beta
_{11}+2(\rvert \alpha _{11}\rvert ^{2}+\rvert \alpha _{12}\rvert ^{2})\beta
_{12}\}\}.  \label{BC}
\end{equation}

\begin{figure}[ht]
\begin{center}
\noindent \includegraphics[width=3.in]{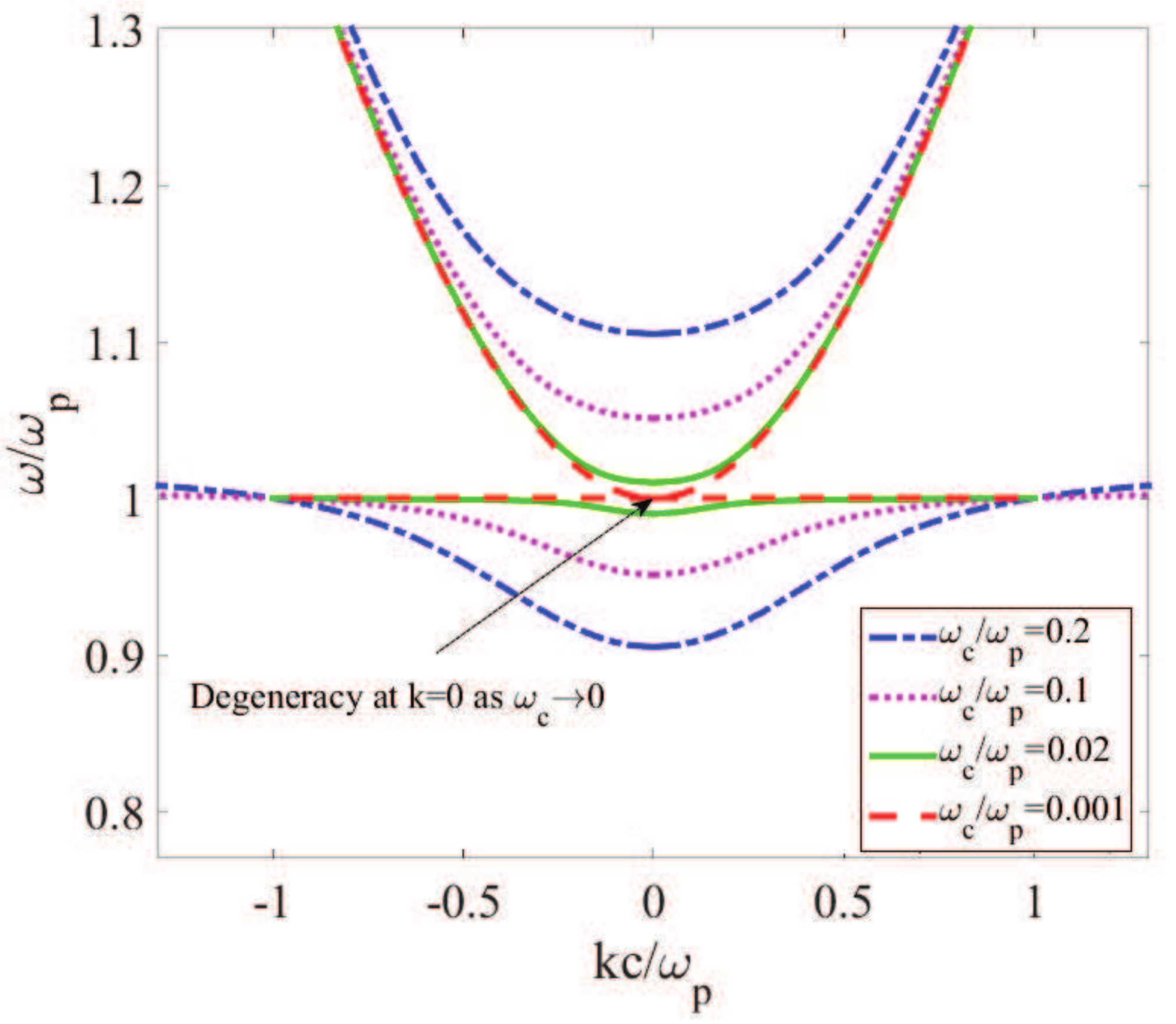}
\end{center}
\caption{Dispersion behavior of bulk TM modes for a magnetoplasma for various values of $\omega_c$. Except for the case $\protect\omega_c=0$, the upper (lower) bands have Chern number +1 (-2).}
\label{Figp1}
\end{figure}

Figure \ref{Figp1} shows the band diagram for TM modes for $\omega_p
/2\pi = 10 $ THz and various values of $\omega_c$. It is clear that as $
\omega_c \rightarrow 0 $ (no magnetic bias, TR symmetry/reciprocity
respected) the two modes become degenerate $k=0, \omega=\omega_p$. As the bias is turned on the material breaks TR
symmetry, $\varepsilon_{12}\neq 0$, and the two modes split apart, creating a bandgap. Note that $\nabla \cdot \mathbf{F}\neq 0$ in general, and so the `monopole' singularity discussed previously is a distributed source for this example, due to the 2D nature of the eigenfunctions.

Details of the Chern number calculation are provided in \cite{Hassani2} and \cite{Hassani1} following the method presented in \cite{Mario2}. Using (\ref{biased_plasma}), the upper band has an integer Chern number, but the lower band does not. As carefully explained in \cite{Mario2}, this is due to ill-behavior of the Hamiltonian for large wavenumber. Integer Chern numbers for both branches are obtained using a non-local material response with a high-wavenumber spatial cutoff,
\begin{equation}
\boldsymbol{\mathrm{M}}_{\text{reg}}(\omega ,\mathbf{k})=\boldsymbol{\mathrm{%
M}}_{\infty }+\frac{1}{1+\mathrm{k}^{2}/\mathrm{k}_{\mathrm{max}}^{2}}%
\left\{ \boldsymbol{\mathrm{M}}({\omega })-\boldsymbol{\mathrm{M}}_{\infty
}\right\}   \label{NLM}
\end{equation}
where $\boldsymbol{%
\mathrm{M}}_{\infty }=\lim\limits_{\omega \rightarrow \infty }\boldsymbol{%
\mathrm{M}}({\omega })$ and the spatial cutoff $\mathrm{k}_{\text{max}}\ $
determines the strength of non-locality such that as $\mathrm{k}_{\text{max}%
}\ \rightarrow \infty $ the material model becomes local. For this
nonlocal material response, the band diagram and Chern numbers have
been obtained using $\mathrm{k}_{\mathrm{max}}=100|\omega _{c}|/c$
(the Chern number calculation is insensitive to the value of
$\mathrm{k}_{\mathrm{max}}$). The resulting Chern numbers for the
two bands are $+1$ for the upper band and $-2$ for the lower band. Note that 
$\sum_{n}C_{n}=0$, since there is also a mode near $\omega=0$, not shown, that has Chern number $1$ \cite{bulkedge}. The Chern numbers are the same for all values
$\omega_c>0$. Figure \ref{Figp1} was obtained using the local model,
but in Fig. \ref{NL} the nonlocal dispersion behavior is shown.

In light of the distinction between the (i) spin-redirection/Rytov--Vladimirskii--Berry) geometric phase and the (ii) Pancharatnam--Berry phase discussed in Section \ref{Intro}, it is worthwhile to note that the spin-redirection phase can be divided into several classes; (a) evolution along one ray as the wave propagates in an inhomogeneous medium, or a curved medium, such as the curved-waveguide example of Fig. 1, and (b) relative phases between different rays (k-vectors) with different directions, such as considered here for the biased plasma. Other examples of case (b) are shown in \cite{Bliokh} and \cite{x} where the Berry phase between different rays results in the spin Hall effect and other spin-orbit phenomena.

\begin{figure}[ht]
\begin{center}
\noindent \includegraphics[width=3.in]{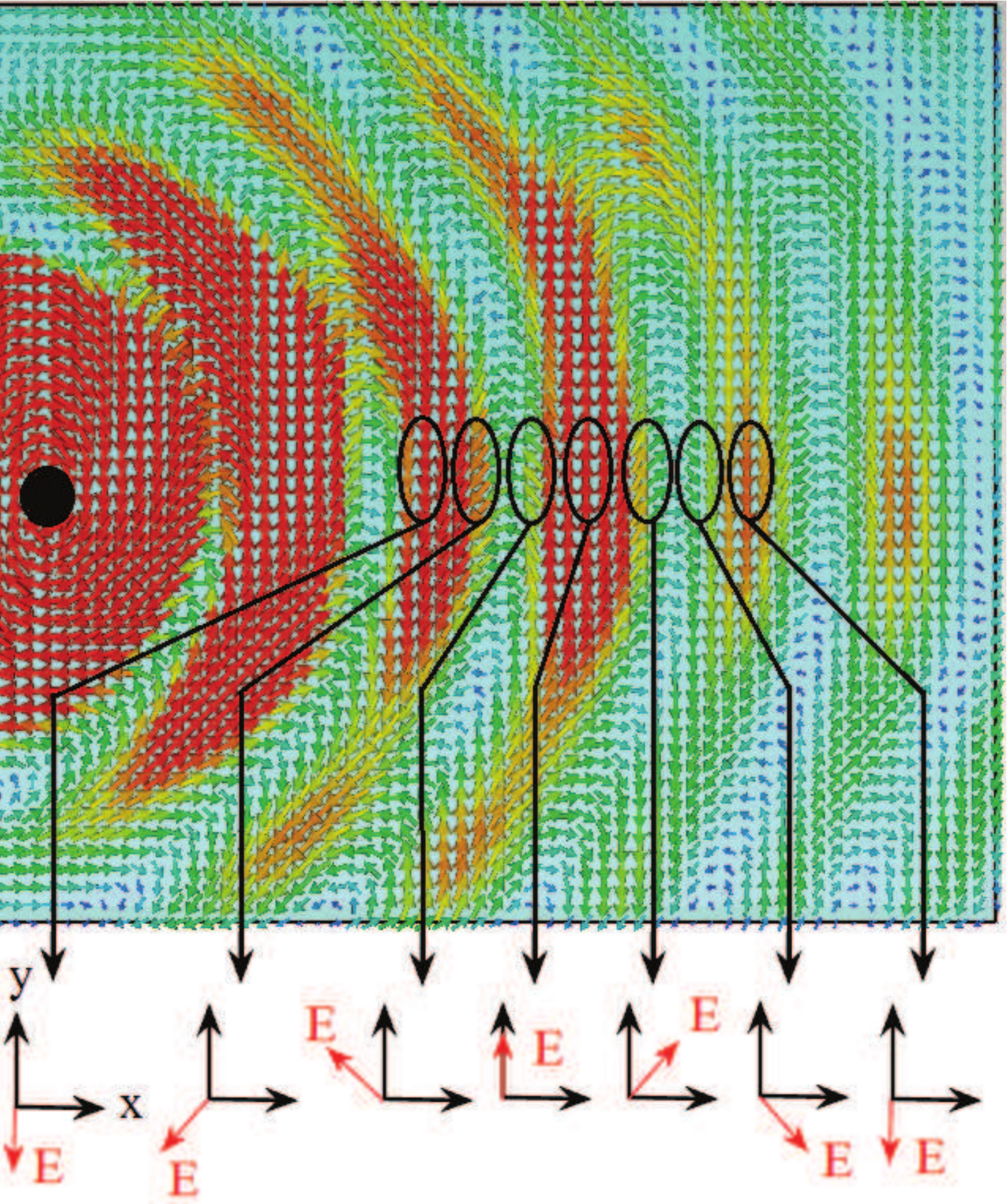}
\end{center}
\caption{Electric field radiated by a horizontal dipole (black circle) in a homogeneous plasma having $\omega/2\pi = 10$ THz, $\omega_p/\omega = 0.84$, and $\omega_c/\omega=0.15$.}
\label{Fig_rot}
\end{figure}

The components of the electric field along the $x$ and $y$ axes in terms of $%
\mathrm{H}_z $ are

\begin{align}  \label{TM_MP}
&\mathrm{E}_x = \frac{ - \varepsilon _{11} \mathrm{k}_y - i\varepsilon _{12} \mathrm{%
k}_x }{ \varepsilon _0 \omega_n (\varepsilon _{11}^2 - \varepsilon _{12}^2 ) } \mathrm{H}%
_z, ~~ \mathrm{E}_y = \frac{ -i \varepsilon _{12} \mathrm{k}_y + \varepsilon _{11}
\mathrm{k}_x }{ \varepsilon _0 \omega_n (\varepsilon _{11}^2 - \varepsilon _{12}^2 ) }
\mathrm{H}_z.
\end{align}
Converting to a polar coordinate system ($r,\varphi$) where $\varphi$ is measured from $\mathbf{k}$,
\begin{equation}
\mathbf{E}=E_{\varphi }\left( \widehat{\mathbf{\varphi }}+\widehat{\mathbf{r}}%
\left( \frac{-i\varepsilon _{12}}{\varepsilon _{11}}\right) \right), \label{Eq:R_C}
\end{equation}%
where $E_{\varphi }=kH_{z}/\varepsilon _{0}\varepsilon _{\text{eff}}\omega _{n}$. It can be seen that there is a quadrature phase relation between the components of the electric field, which will produce a rotation of the electric field in the plane of wave propagation ($x-y$). Generally, an elliptical polarization is produced due to non-equal $\varepsilon _{11} $ and $\varepsilon _{12}$. The instantaneous electric field is
\begin{equation}
\boldsymbol{\mathrm{E}}=\mathrm{E}_{\varphi }\left( \mathrm{cos}(\omega _{n}t-%
\mathrm{k}r)\hat{\mathbf{\varphi }}+\frac{\varepsilon _{12}}{\varepsilon _{11}}%
\mathrm{sin}(\omega _{n}t-\mathrm{k}r)\hat{\mathbf{r}}\right) 
\end{equation}
where $r=\sqrt{x^{2}+y^{2}}$. Figure \ref{Fig_rot} shows the electric field vector due to a horizontal source in a biased magnetoplasma at a fixed instant of time, computed using CST Microwave Studio, for $\omega/2\pi=10 \text{ THz}, \omega_p/\omega=0.84, \text{and } \omega_c/\omega=0.15$. The ellipses span one wavelength, and highlight how the field rotates as the wave propagates (at a fixed point in space, the field also rotates as time progresses). If the bias is turned off ($\varepsilon _{12} = 0 $), the material is reciprocal ($\varepsilon_{12}=0$), the longitudinal electric field is zero so that the field does not rotate as described above, the eigenfunctions are real-valued, and all Berry quantities vanish ($\mathbf{A}=\mathbf{F}=\mathbf{0}$). 

The Berry connection and phase are defined solely in terms of the envelope of the eigenmodes, (\ref{TM_MP}); the propagation factor $e^{i \mathbf{k}\cdot\mathbf{r}}$ is not involved in computing the Berry phase. As described in Sec. \ref{SecMax}, the incremental Berry phase is the relative phase difference between an eigenfunction at $\mathbf{k}$ and at a nearby point $\mathbf{k}+d\mathbf{k}$. Therefore, consistent with the operation $\nabla_{\mathbf{k}}$ used in computing $\mathbf{A}$, we consider a small change in $\mathbf{k}$ to observe the Berry phase. Since the bulk dispersion behavior is isotropic in the $(k_{x}, k_{y})$ plane, to change the value of $\mathbf{k}$ we need to move along an azimuthal arc at fixed radial wavenumber. 

Defining the incremental Berry phase as $\delta \gamma =\mathbf{A}\cdot kd%
\widehat{\mathbf{\varphi}}=A_{\varphi }k\delta \varphi $, and writing $\mathbf{E}=%
\mathbf{E}\left( r,\varphi ,t\right) $, we can form a quantity 
\begin{equation}
Q\left( t\right) =\frac{\mathbf{E}\left( r,0,0\right) \cdot \mathbf{E}\left(
r,\delta \varphi ,t\right) }{\left\vert \mathbf{E}\left( r,0,0\right) \cdot 
\mathbf{E}\left( r,\delta \varphi ,t\right) \right\vert },
\end{equation}%
where $r$ is an arbitrary fixed far-field distance, that measures the similarity between the electric field at $\left( t,\varphi
\right) =\left( 0,0\right) $ and the field at $\left( t,\varphi \right) =\left(
t,\delta \varphi \right) $; when the fields are the same, $Q\left( t\right) =1$%
. For a given small change in $\mathbf{k}$ represented by a small change in
the angle $\delta \varphi $, we expect $Q\left( t\right) $ to be maximized when 
$\omega t=\delta \gamma $, indicating that the incremental Berry phase $%
\delta \gamma $ leads to the correct time shift between nearby eigenfunctions. 

\begin{figure}[h!]
\begin{center}
\noindent \includegraphics[width=3.5in]{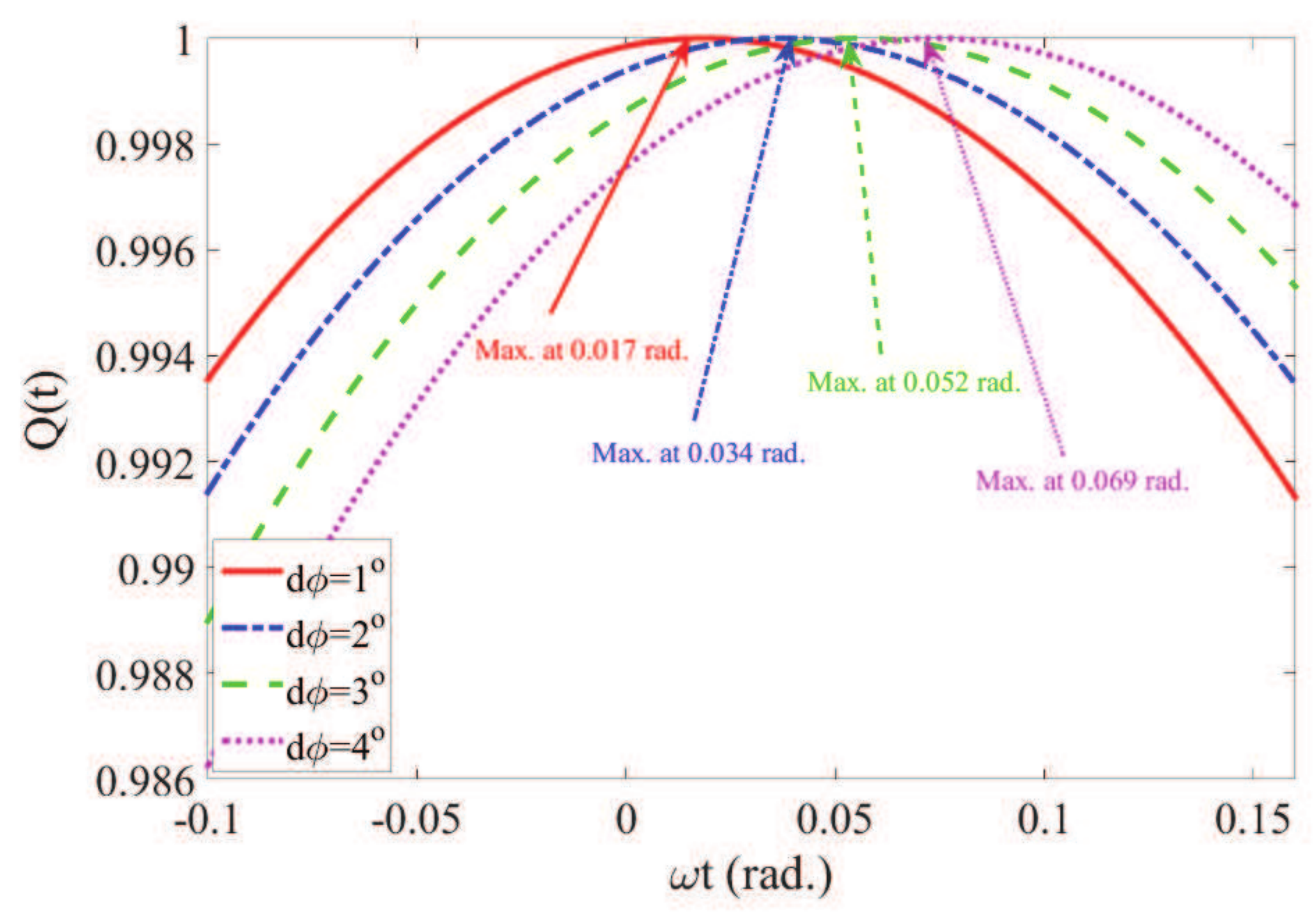}
\end{center}
\caption{$Q(t)$ versus $\omega t$ for four different momentum angles $ \delta \varphi_n = n\pi/180 $ radians, $n=1,2,3,4$, for $ \omega / 2\pi = 10  $ THz, $ \omega_c/2\pi = 1.73  $ THz, and $ \omega_p/2\pi = 9 $ THz}
\label{FigE1}
\end{figure}

As a numerical example, for $ \omega / 2\pi = 10  $ THz, $ \omega_c/2\pi = 1.73  $ THz, and $ \omega_p/2\pi = 9 $ THz, we have bulk propagation since $ \varepsilon _{eff}>0 $. We consider four small angles, $ \delta \varphi_n = n\pi/180 $ radians (i.e., $n^o$), $n=1,2,3,4$. For these values of $\delta\varphi_n$ the Berry phases are $\delta \gamma_1 = 0.017 $ rad, $\delta \gamma_2 = 0.034$ rad, $\delta \gamma_3 = 0.052 $ and $\delta \gamma_4 = 0.069 $ rad. respectively. Figure \ref{FigE1} indeed shows that for each angle, the Berry phase leads to the correct time shift between eigenfunctions separated by $\delta k$.

Furthermore, it should be noted that we can have Berry phase with no field rotation. For example, consider an antenna that radiates a linearly polarized plane wave that propagates along $z$. The antenna radiates different frequencies (in time, the fields form a traveling pulse). Assume that the amplitude of all harmonics is the same, but that the phase differs. Depending on the relative phase between harmonics, one can have a nontrivial Berry connection $\mathrm{A}_z$ because $\mathbf{A}$ is gauge dependent, and the relative
phase (between one $k$ value, associated with one frequency, and a nearby $k$ value, associated with a different frequency) effectively changes the gauge. Therefore, one can have nontrivial Berry quantities in realistic physical scenarios even with linear polarizations, due to the gauge dependence (changing the phase of an eigenmode is a gauge change), and because the Berry connection along a single (non-closed) line does not determine an invariant.

\subsection{Edge Plasmons}

Surface plasmon polaritons (SPPs) are electromagnetic modes
localized near the boundary between two materials, typically a
material having negative permittivity and one having positive
permittivity. For simple materials, these edge modes can travel in
any direction along the edge. If, however, the edge mode is sufficiently non-reciprocal, it can be unidirectional. This was
realized long ago for biased plasma interfaces \cite{SR}-\cite{ZR};
here we recognize the effect to be subsumed with the Berry
phase/Berry curvature framework and to the
topological properties of the involved materials.

The occurrence of gap Chern number $\mathcal{C}_{\text{gap},\Delta
}=1$ predicts the presence of
a single, unidirectional surface mode. This can be confirmed by
directly solving Maxwell's equation for surface modes at the
interface of a biased plasma and a trivial medium with permittivity
$ \varepsilon _s $. Assume fields invariant along $z$ and propagating
along $x$, so that in the dielectric ($y>0$) we have
field variation $e^{i\mathrm{k}_{\mathrm{spp}} x}e^{-\alpha _{s}y}$, and in the plasma ($y<0$), $%
e^{i\mathrm{k}_{\mathrm{spp}} x}e^{\alpha _{p}y}$. The condition
$\alpha_p, \alpha_s>0$ define the proper Riemann sheets for the
wavenumber. Plugging into Maxwell's equations leads to the TM SPP
fields

\begin{align}
\left(
\begin{array}{c}
E_{x} \\
E_{y} \\
H_{z}%
\end{array}%
\right) & =\left(
\begin{array}{c}
E_{0x} \\
E_{0y} \\
H_{0z}%
\end{array}%
\right) e^{i\mathrm{k}_{\mathrm{spp}}x}\left\{
\begin{array}{c}
e^{-\alpha _{s}y},\ \ y>0 \\
e^{\alpha _{p}y},\ \ \ y<0%
\end{array}%
\right.  \\
E_{0x}& =\frac{i\alpha _{s}}{\omega \varepsilon _{0}\varepsilon _{s}}%
H_{0z},\ \ E_{0y}^{s}=\frac{-\mathrm{k}_{\mathrm{spp}}}{\omega \varepsilon
_{0}\varepsilon _{s}}H_{0z} \\
E_{0y}^{p}& =\frac{1}{\omega \varepsilon _{0}}\frac{\varepsilon _{12}\alpha
_{p}-\mathrm{k}_{\mathrm{spp}}\varepsilon _{11}}{\varepsilon
_{11}^{2}-\varepsilon _{12}^{2}}H_{0z}
\end{align}
where $k_{s}^{2}=\omega ^{2}\mu \varepsilon _{0}\varepsilon _{s}$, with the SPP dispersion equation \cite{Arthur,Hassani3}
\begin{equation}\label{Eq:SPP}
\frac{\alpha_s}{\varepsilon _s} + \frac{\alpha_p}{\varepsilon _{eff}} = \varepsilon _{12} \frac{\mathrm{k}_{\mathrm{spp}}}{\varepsilon _{11} \varepsilon _{eff}}
\end{equation}
where $\mathrm{k}_{\mathrm{spp}}$ is the propagation constant of the surface plasmon-polariton (SPP) along the interface, $ \alpha_s = k_0 \sqrt{ \left( \mathrm{k}_{\mathrm{spp}} / k_0 \right)^2  - \varepsilon _{s}} $ and $ \alpha_p = k_0 \sqrt{ \left( \mathrm{k}_{\mathrm{spp}} / k_0 \right)^2  - \varepsilon _{eff}} $ and $ \varepsilon _{eff} = (\varepsilon ^2_{11} - \varepsilon ^2_{12}) / \varepsilon _{11} $ are the attenuation constants in the simple and gyro-electric media, respectively. For the general case, (\ref{Eq:SPP}) cannot be solved analytically.

\subsubsection{Limit $|\varepsilon_s|\rightarrow \infty$}
In the limit that $|\varepsilon_s|\rightarrow \infty$ (implementing a perfect conductor), the dispersion equation can
 be solved to yield $\mathrm{k}_{\mathrm{spp}}=k_0 \sqrt{\varepsilon_{11}}$, such that $\alpha _{p}=k_{0}\varepsilon _{12}/\sqrt{\varepsilon _{11}}$. For this case, $E_{0x}\rightarrow 0$ and the generally TM SPP becomes a TEM mode \cite{SR}. It is easy to see that we require $\varepsilon_{11}>0$ to have a solution, and so we need to operate such that $\omega>\omega_p$. For $\varepsilon _{12}>0$ there are no solutions $\mathrm{k}_{\mathrm{spp}}<0$, since if we take the negative root of $\sqrt{\varepsilon _{11}}$ then $\alpha_p<0$ and the mode exponentially increases away from the interface (i.e., it is on the wrong Riemann sheet). Similarly, if we have $\varepsilon _{12}<0$ there are no solutions $\mathrm{k}_{\mathrm{spp}}>0$. Therefore, we have a unidirectional SPP. For the unbiased case ($\varepsilon _{12}=0$), there is no SPP solution.

\subsubsection{$|\varepsilon_s|$ Finite}
For finite values of $\varepsilon_s<0$, an SPP is also obtained for $\omega>\omega_p$, where $\varepsilon_{11}>0$. For $\varepsilon_s>0$, an SPP occurs for $\omega<\omega_p$, but this case is not of interest since at a discontinuity radiation into the upper bulk region would occur.

\bigskip

\begin{figure}[ht]
    \begin{center}
        \noindent  \includegraphics[width=3.in]{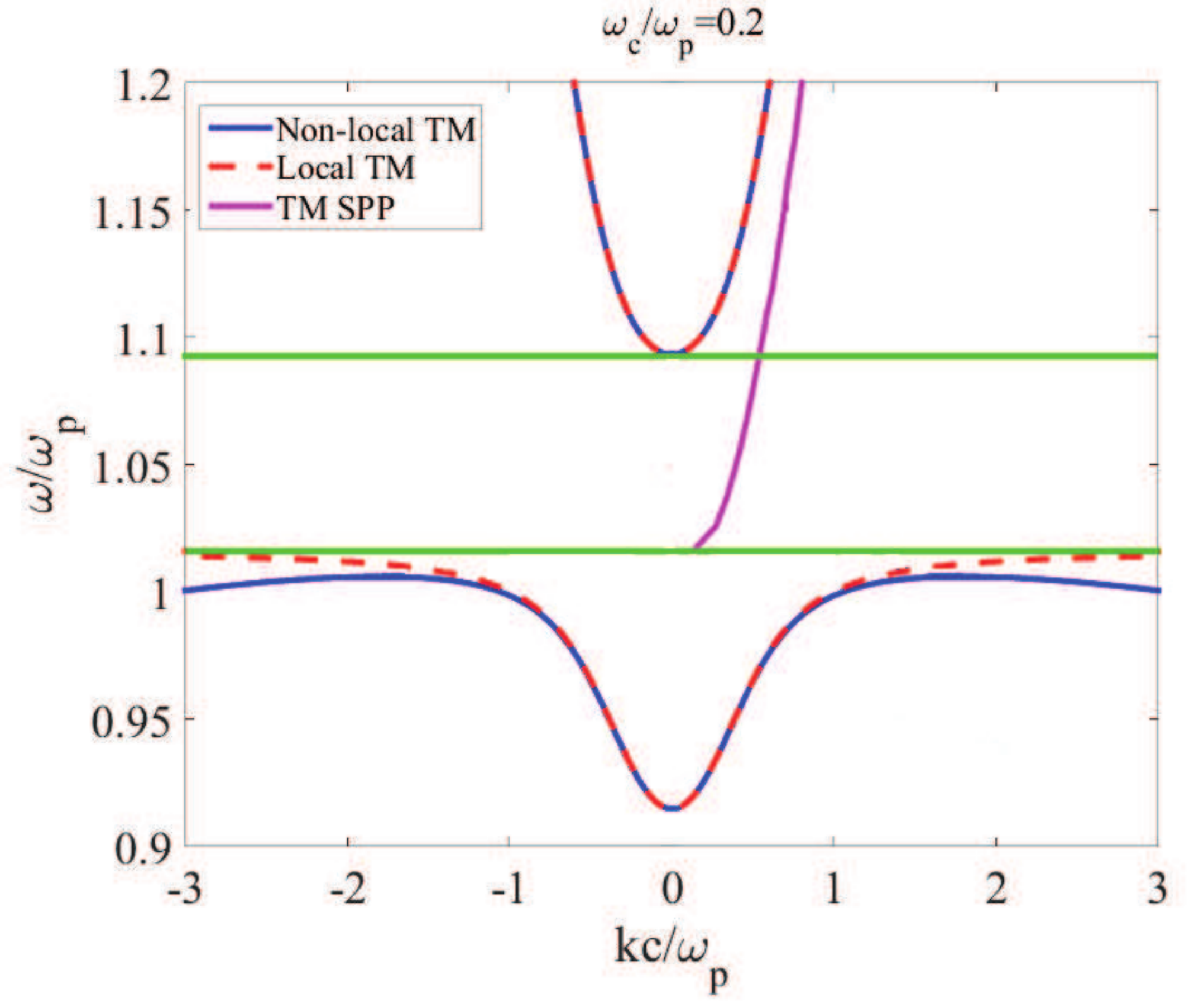}
    \end{center}
    \caption{Dispersion behavior of bulk (blue and dashed red) and SPP (purple) modes for the interface between perfect conductor and biased non local plasma. Biased plasma has $\omega_c/\omega_p=0.2 $ and for non local case $k_{\text{max}}=100 \omega_c / c$.}
    \label{NL}
\end{figure}

\begin{figure}[ht]
    \begin{center}
        \noindent  \includegraphics[width=3.in]{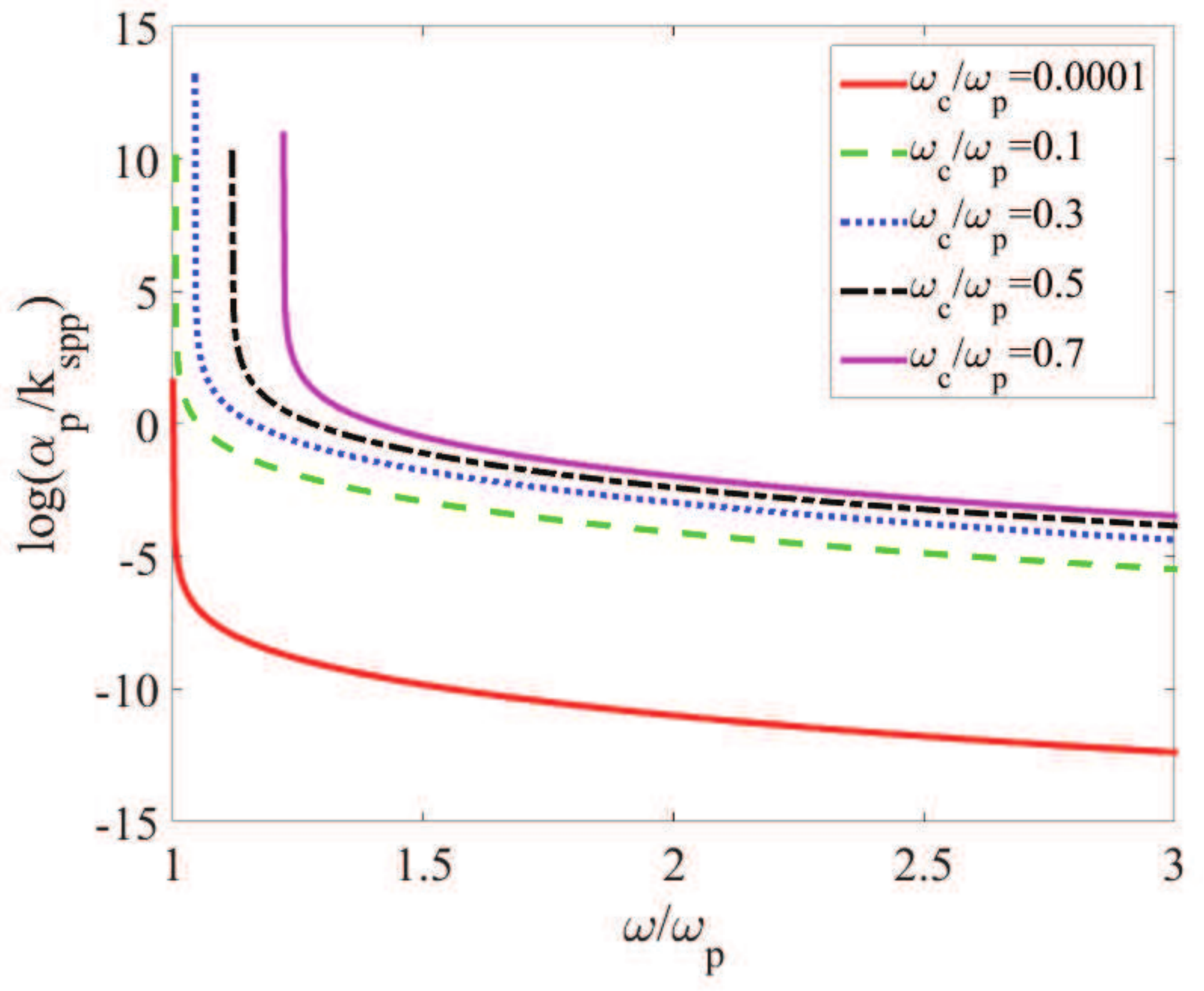}
    \end{center}
    \caption{Confinement factor for SPP mode for the interface between a plasma having $\omega_p/\omega=0.97$ and a perfect conductor.}
    \label{a_p}
\end{figure}

\begin{figure}[ht]
    \begin{center}
        \noindent  \includegraphics[width=2.5in]{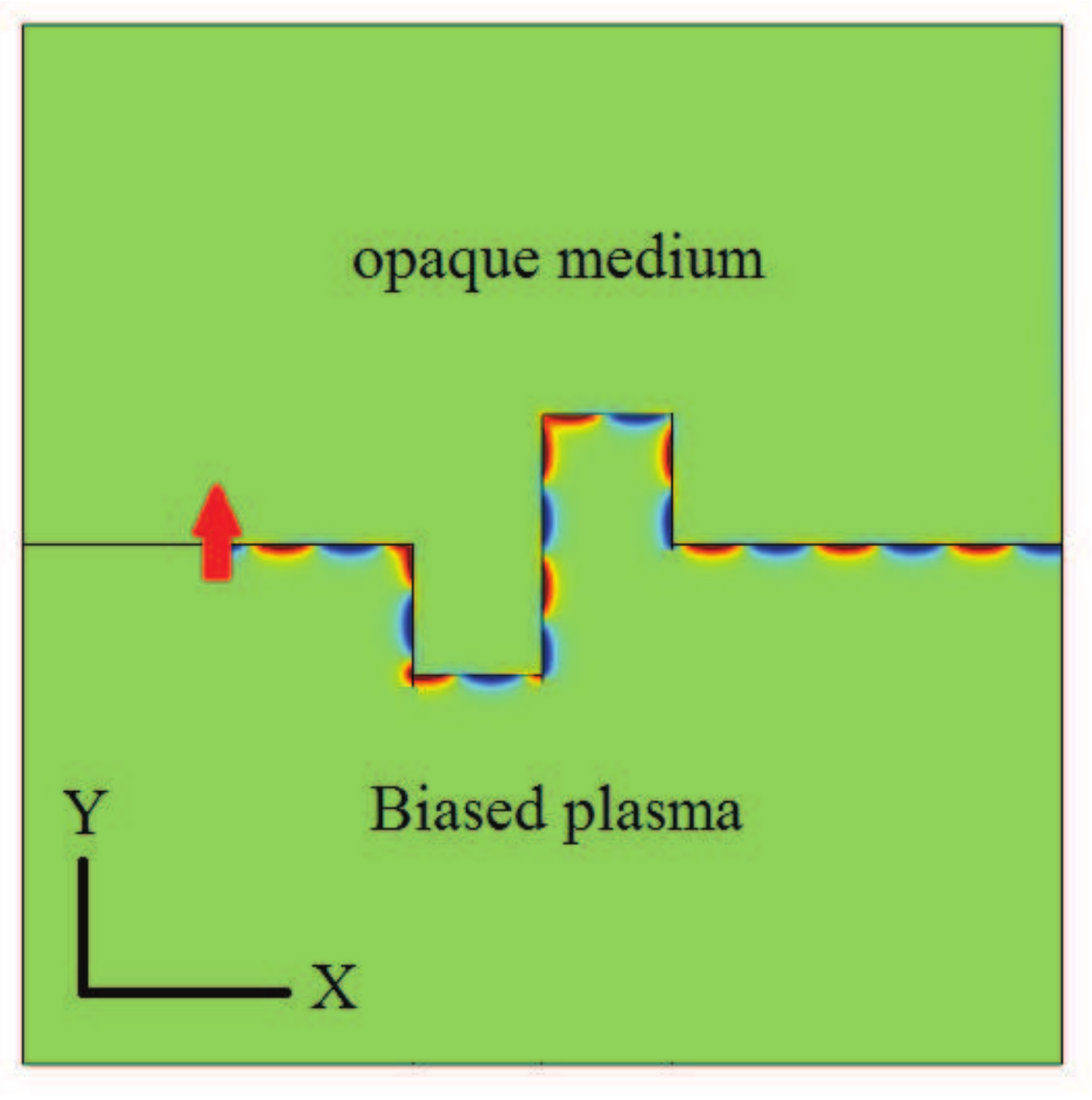}
    \end{center}
    \caption{SPP excited by a vertical 2D source at the interface between a biased plasma having $\omega/2\pi = 10$ THz, $\omega_p/\omega = 0.97$, and $\omega_c/\omega=0.173$ and an opaque material having $\varepsilon_s=-2$.}
    \label{SPP}
\end{figure}

Figure \ref{NL} shows the bulk modes (blue and dashed red) for a biased
plasma having $\omega_c/\omega_p=0.2$, for both the local and
nonlocal (\ref{NLM}) models. Also shown is the SPP dispersion
(purple) for the interface between the plasma and a perfect
conductor. The SPP line crosses the gap (denoted by green lines) of the magnetoplasma with
monotonic slope ($ \partial_{\mathrm{k}} \omega = v_g >0 $). Fig.
\ref{a_p} shows the SPP confinement factor $\alpha_p$ as a function
of frequency and bias, also for the magnetoplasma-PEC interface. As
the bias is increased, the SPP becomes more confined to the
interface, and as frequency increases the mode becomes less
well-confined. Maximum confinement occurs at
$\omega=\left({\omega_p}^2 + {\omega_c}^2 \right )^{1/2}$. Figure \ref{SPP} shows the electric field distribution at an interface
between a biased plasma and a medium having $\varepsilon_s=-2$.
Although the interface has several sharp discontinuities, since the
SPP is unidirectional it cannot backscatter, and since we operate
in the bandgap of the plasma, there can be no diffraction/radiation
into the bulk (the $\varepsilon_s=-2$ region is opaque).

This example nicely illustrates the implications of the bulk-edge
correspondence principle to topological continua. In this case,  the
bulk-edge correspondence works even disregarding the impact of the
spatial cut-off in the waveguiding problem. In general, such effects
cannot be neglected and the high-frequency cut-off needs to be mimicked
in the realistic physical scenario by introducing a small air-gap in
between the two materials \cite{Mario2,bulkedge}.

\section{Conclusion}

The most celebrated property of photonic topological insulators (PTIs) is their capability to support one-way surface plasmon-polaritons at an interface that are immune to backscattering from defects or imperfections. These effects were first developed at the electronic level in quantum mechanics, by systems evolving in time in a cyclic fashion. In this paper, we have demonstrated that all Berry quantities (Berry phase, connection/potential, and curvature) can be analytically obtained and interpreted from a fully classical electromagnetic perspective. We have discussed the physical meaning of the Berry phase, connection, and curvature, how these quantities arise in electromagnetic problems, and the significance of Chern numbers for unidirectional, scattering-immune SPP propagation.
\\

\noindent{Acknowledgments}
\\

\noindent We would like to thank Konstantin Bliokh for considerable help in improving the manuscript.

\section{Appendix}

Here we show how time-reversal and inversion symmetry affects the Berry
connection and curvature, and that for a loss-free medium, a reciprocal
medium is a medium with time-reversal symmetry and vice versa.

\subsubsection{Time-reversal}

Time reversal transformation (e.g., changing $t\rightarrow -t$) for
electromagnetic quantities arise from the fact that charges move in opposite
directions under TR, which leaves the electric field unchanged but changes
the direction of the magnetic field. Maxwell's equations are invariant under
TR, and we have \cite{Altman}, \cite{TL}.
\begin{align}
& \mathcal{T~}\boldsymbol{\mathrm{E}}\left( \mathbf{r},t\right) =\boldsymbol{%
\mathrm{E}}\left( \mathbf{r},-t\right) ,\ \mathcal{T~}\mathbf{D}\left(
\mathbf{r},t\right) =\mathbf{D}\left( \mathbf{r},-t\right)  \\ \notag
& \mathcal{T~}\mathbf{B}\left( \mathbf{r},t\right) =-\mathbf{B}\left(
\mathbf{r},-t\right) ,\ \mathcal{T~}\mathbf{H}\left( \mathbf{r},t\right) =-%
\mathbf{H}\left( \mathbf{r},-t\right)  \\ \notag
& \mathcal{T~}\mathbf{J}\left( \mathbf{r},t\right) =-\mathbf{J}\left(
\mathbf{r},-t\right) ,\ \mathcal{T~}\rho \left( \mathbf{r},t\right) =\rho
\left( \mathbf{r},-t\right) \ \ \ \ \ \ \ ~~
\end{align}%
where $\mathcal{T}$ is the time-reversal operator.

To see the effect of TR on momentum-frequency domain quantities, taking the
Fourier transform of the time-reversed field $\mathbf{E}\left( \mathbf{r}%
,-t\right) $ leads to
\begin{align}
\mathcal{T\ }\mathbf{E}\left( \mathbf{k},\omega \right) & \equiv
\int\nolimits_{\mathbf{r}}\int\nolimits_{t=-\infty }^{t=\infty }\mathcal{T~%
}\mathbf{E}\left( \mathbf{r},t\right) e^{i\mathbf{k}\cdot \mathbf{r}%
}e^{-i\omega t}d\mathbf{r}dt\\ \notag
&=\int\nolimits_{\mathbf{r}}\int\nolimits_{t=-%
\infty }^{t=\infty }\mathbf{E}\left( \mathbf{r},-t\right) e^{i\mathbf{k}%
\cdot \mathbf{r}}e^{-i\omega t}d\mathbf{r}dt \\ \notag
& =-\int\nolimits_{\mathbf{r}}\int\nolimits_{-\tau =-\infty }^{-\tau
=\infty }\mathbf{E}\left( \mathbf{r},\tau \right) e^{i\mathbf{k}\cdot
\mathbf{r}}e^{i\omega t}d\mathbf{r}d\tau \\ \notag
&=\int\nolimits_{\mathbf{r}%
}\int\nolimits_{\tau =-\infty }^{\tau =\infty }\mathbf{E}\left( \mathbf{r}%
,\tau \right) e^{i\mathbf{k}\cdot \mathbf{r}}e^{i\omega t}d\mathbf{r}d\tau
\\ \notag
& =\mathbf{E}\left( \mathbf{k},-\omega \right) =\mathbf{E}^{\ast }\left( -%
\mathbf{k},\omega \right) ,
\end{align}%
where we note that applying $\mathcal{T}$ to a quantity that is not a
function of time, e.g., $\mathcal{T}\mathbf{E}\left( \mathbf{k},\omega
\right) $, means applying $\mathcal{T}$ in the time-domain and then taking
the appropriate transform. Therefore, the prescription is that applying TR in
the time domain is equivalent to taking complex conjugate and reversing
momentum (or, not taking complex conjugate and reversing $\omega \,$). For
the magnetic field, we get $\mathcal{T}\mathbf{H}\left( \mathbf{k},\omega
\right) =-\mathbf{H}^{\ast }\left( -\mathbf{k},\omega \right) $.

For a source-driven problem, $\mathbf{k}$ and $\omega$ are independent parameters. However, for a
source-free (eigenmode) problem, $\mathbf{k}=\mathbf{k}\left( \omega \right)
$, and the momentum and frequency variables are linked by a dispersion
equation (and for a trivial dispersion equation like $k=\omega \sqrt{\mu
\varepsilon }$, reversing $k$ and reversing $\omega $ are equivalent). In
this case,%
\begin{align}
\mathcal{T\ }\mathbf{E}\left( \mathbf{k},\omega \right) & \equiv
\int\nolimits_{\mathbf{r}}\int\nolimits_{t=-\infty }^{t=\infty }\mathcal{T~%
}\mathbf{E}\left( \mathbf{r},t\right) e^{i\mathbf{k}\left( \omega \right)
\cdot \mathbf{r}}e^{-i\omega t}d\mathbf{r}dt\\ \notag
&=\int\nolimits_{\mathbf{r}%
}\int\nolimits_{t=-\infty }^{t=\infty }\mathbf{E}\left( \mathbf{r}%
,-t\right) e^{i\mathbf{k}\left( \omega \right) \cdot \mathbf{r}}e^{-i\omega
t}d\mathbf{r}dt \\ \notag
& =-\int\nolimits_{\mathbf{r}}\int\nolimits_{-\tau =-\infty }^{-\tau
=\infty }\mathbf{E}\left( \mathbf{r},\tau \right) e^{i\mathbf{k}\left(
\omega \right) \cdot \mathbf{r}}e^{i\omega \tau}d\mathbf{r}d\tau \\ \notag
&=\int\nolimits_{\mathbf{r}}\int\nolimits_{\tau =-\infty }^{\tau =\infty }%
\mathbf{E}\left( \mathbf{r},\tau \right) e^{i\mathbf{k}\left( \omega \right)
\cdot \mathbf{r}}e^{i\omega \tau}d\mathbf{r}d\tau  \\ \notag
& =\mathbf{E}^{\ast }\left( -\mathbf{k},\omega \right)  \\ \notag
& =\binom{\mathbf{E}\left( \mathbf{k},-\omega \right) \text{ \ \ if }\mathbf{%
k}\left( \omega \right)=\mathbf{k^{\text {even}}}\left( \omega \right) }{\mathbf{E}%
\left( -\mathbf{k},-\omega \right) \text{ \ \ if }\mathbf{k}\left( \omega
\right) =\mathbf{k^{\text {odd}}}\left( \omega \right) }.
\end{align}%
Therefore,
\begin{align}
\mathcal{T\ }\boldsymbol{f}_{n,\mathbf{k}\left( \omega \right) }& =\left(
\begin{array}{c}
\boldsymbol{\mathrm{E}}^{\ast }\left( -\mathbf{k}\left( \omega \right)
\right)  \\
-\boldsymbol{\mathrm{H}}^{\ast }\left( -\mathbf{k}\left( \omega \right)
\right)
\end{array}%
\right) =\,\,\boldsymbol{\mathrm{T}}_{6\times 6}\cdot \boldsymbol{f}_{n,-%
\mathbf{k}\left( \omega \right) }^{\ast }  \notag \\
& =\,\,\binom{\boldsymbol{\mathrm{T}}_{6\times 6}\cdot \boldsymbol{f}_{n,%
\mathbf{k}\left( -\omega \right) }\text{ \ if }\mathbf{k}\left( \omega
\right) =\mathbf{k}^{\text{even}}\left( \omega \right) \text{ }}{\boldsymbol{%
\mathrm{T}}_{6\times 6}\cdot \boldsymbol{f}_{n,-\mathbf{k}\left( -\omega
\right) }\text{ \ if }\mathbf{k}\left( \omega \right) =\mathbf{k}^{\text{odd}%
}\left( \omega \right) \text{.}}  \label{TEF}
\end{align}
where
\begin{equation}
\boldsymbol{\mathrm{T}}_{6\times 6}=\left(
\begin{array}{cc}
\boldsymbol{\mathrm{I}}_{3\times 3} & \boldsymbol{\mathrm{0}} \\
0 & -\boldsymbol{\mathrm{I}}_{3\times 3}%
\end{array}%
\right)
\end{equation}%
is the Poynting vector reversing operator (applying this operator to $\boldsymbol{f}_{n}$ reverses the
direction of the group velocity) \cite{Altman}, such that $\boldsymbol{%
\mathrm{T}}_{6\times 6}\cdot \boldsymbol{\mathrm{T}}_{6\times 6}=\boldsymbol{%
\mathrm{I}}_{6\times 6}$. Note that $\boldsymbol{\mathrm{T}}_{6\times 6}=\mathbf{\sigma }_{z}$, the $%
6\text{x}6$ form of the Pauli spin matrix. If the system is $\mathcal{T}$-invariant, $\boldsymbol{%
\mathrm{T}}_{6\times 6}\cdot \boldsymbol{f}_{n,-\mathbf{k}\left( \omega
\right) }^{\ast }=$ $e^{i\zeta \left( \mathbf{k}\right) }\boldsymbol{f}_{n,%
\mathbf{k}\left( \omega \right) }$, where we can include an arbitrary phase
since eigenfunctions are defined up to a phase factor. As an example, for a simple plane wave in vacuum (which is $\mathcal{T}$%
-invariant), let $\mathbf{E}=\widehat{\mathbf{x}}E_{0}e^{ik_{0}z}$, where $%
k_{0}\left( \omega \right) =\omega \sqrt{\varepsilon _{0}\mu _{0}}$. Then, $%
\mathbf{H}=\widehat{\mathbf{y}}E_{0}\left( k_{0}/\omega \mu _{0}\right)
e^{ikz}$, and $\boldsymbol{f}_{n,\mathbf{k}\left( \omega \right) }=\left(
e^{ik_{0}z},0,0,0,\frac{k_{0}}{\omega \mu _{0}}e^{ik_{0}z},0\right) ^{\text{T%
}}$, and it is easy to see that $\mathcal{T\ }\boldsymbol{f}_{n,\mathbf{k}%
\left( \omega \right) }=\boldsymbol{\mathrm{T}}_{6\times 6}\cdot \boldsymbol{%
f}_{n,-\mathbf{k}\left( \omega \right) }^{\ast }=\boldsymbol{f}_{n,\mathbf{k}%
\left( \omega \right) }$.

Starting with (\ref{KD}) we can obtain a TR eigenfunction $\mathcal{T~}\mathbf{w}_{n,\mathbf{k}\left(
\omega \right) }$ by considering that in the time domain $\mathbf{w}$ is a
convolution of the inverse temporal transforms of the two terms. Time reversal of each term leads to, by the convolution theorem, the product
of the individual time-reversed temporal transforms. Using (\ref{TR1}),
\begin{align}
& \mathcal{T}\mathbf{w}_{n,\mathbf{k}\left( \omega \right) }  \notag \\
& =\boldsymbol{\mathrm{T}}_{6\times 6}\cdot \left[ \partial _{\omega }\left(
\omega \boldsymbol{\mathrm{M}}^{\ast }\left( \omega ,-\mathbf{k}\right)
\right) \right] ^{1/2}\cdot \boldsymbol{\mathrm{T}}_{6\times 6}\cdot
\boldsymbol{\mathrm{T}}_{6\times 6}\cdot \boldsymbol{f}_{n,-\mathbf{k}\left(
\omega \right) }^{\ast }  \notag \\
& =\boldsymbol{\mathrm{T}}_{6\times 6}\cdot \mathbf{w}_{n,-\mathbf{k}\left(
\omega \right) }^{\ast }.
\end{align}
Then, it can be shown that%
\begin{equation}
\mathcal{T~}\boldsymbol{\mathrm{A}}_{n}\left( \mathbf{k}\right) =\boldsymbol{%
\mathrm{A}}_{n}(-\mathbf{k})  \label{tr11}
\end{equation}%
and that, if a system is $\mathcal{T}$-invariant,
\begin{equation}
\boldsymbol{\mathrm{A}}_{n}(-\mathbf{k})=\boldsymbol{\mathrm{A}}_{n}(\mathbf{%
k})+\nabla _{\mathbf{k}}\xi \left( \mathbf{k}\right)   \label{tr12}
\end{equation}%
where $\xi \left( \textbf{k}\right)$  is an arbitrary phase. To see
this, consider that
\begin{align}
\mathcal{T\ }\boldsymbol{\mathrm{A}}_{n}(\mathbf{k})& =i\left( \mathcal{T~}%
\mathbf{w}_{n,\mathbf{k}}\right) ^{\dagger }\cdot \nabla _{\mathbf{k}}\left(
\mathcal{T~}\mathbf{w}_{n,\mathbf{k}}\right)   \notag \\
& =i\,\,\left( \boldsymbol{\mathrm{T}}_{6\times 6}\cdot \mathbf{w}_{n,-%
\mathbf{k}}^{\ast }\right) ^{\dagger }\cdot \nabla _{\mathbf{k}}\left(
\boldsymbol{\mathrm{T}}_{6\times 6}\cdot \mathbf{w}_{n,-\mathbf{k}}^{\ast
}\right)   \notag \\
& =i\,\,\mathbf{w}_{n,-\mathbf{k}}^{\text{T}}\cdot \nabla _{\mathbf{k}}%
\mathbf{w}_{n,-\mathbf{k}}^{\ast }  \notag \\
& =-i\left( \mathbf{w}_{n,-\mathbf{k}}^{\dag }\cdot \nabla _{-\mathbf{k}}%
\mathbf{w}_{n,-\mathbf{k}}\right) =\mathbf{A}_{n}\left( -\mathbf{k}\right)
\end{align}
since $\mathbf{w}_{n,\mathbf{k}}^{\dagger }\cdot \nabla _{\mathbf{k}}%
\mathbf{w}_{n,\mathbf{k}}$ is pure-imaginary, proving (\ref{tr11}). To prove
(\ref{tr12}), note that if a system is $\mathcal{T}$-invariant,
\begin{equation}
\mathcal{T\ }\mathbf{w}_{n,\mathbf{k}}\left( \mathbf{k}\right) =e^{i\zeta
\left( \mathbf{k}\right) }\, \mathbf{%
w}_{n,\mathbf{k}}
\end{equation}%
from which (\ref{tr12}) follows using the same arguments as in (\ref{gauge}).

From this, for the Berry curvature,
\begin{align}
\mathcal{T~}\boldsymbol{\mathrm{F}}_{n}(\mathbf{k})& =\nabla _{\mathbf{k}%
}\times \left( \mathcal{T~}\mathbf{A}_{n}\left( \mathbf{k}\right) \right)
\nonumber \\
& =\nabla _{\mathbf{k}}\times \left( \mathbf{A}_{n}\left( -\mathbf{k)}%
\right) \right)   \nonumber \\
& =-\nabla _{-\mathbf{k}}\times \left( \mathbf{A}_{n}\left( -\mathbf{k)}%
\right) \right)   \nonumber \\
& =-\boldsymbol{\mathrm{F}}_{n}(-\mathbf{k})
\end{align}
Thus, for a system that respects TR, one can have nonzero Berry curvature,
but, from (\ref{Eq:chern}) the Chern number will be zero ($\mathcal{T}~C=-C$, such that if
TR symmetry is respected, $C=-C$ and so $C=0$).

\subsubsection{Inversion}

\bigskip Inversion (parity) transformation means replacing $\mathbf{r}$ with
$-\mathbf{r}$. Under the inversion symmetry operator $\mathcal{I}$, \cite{TL}
\begin{align}
& \mathcal{I~}\mathbf{E}\left( \mathbf{r},t\right) =-\mathbf{E}\left( -%
\mathbf{r},t\right) ,\ \ \mathcal{I~}\mathbf{D}\left( \mathbf{r},t\right) =-%
\mathbf{D}\left( -\mathbf{r},t\right)  \\ \notag
& \mathcal{I~}\mathbf{B}\left( \mathbf{r},t\right) =\mathbf{B}\left( -%
\mathbf{r},t\right) ,\ \ \mathcal{I~}\mathbf{H}\left( \mathbf{r},t\right) =%
\mathbf{H}\left( -\mathbf{r},t\right)  \\ \notag
& \mathcal{I~}\mathbf{J}\left( \mathbf{r},t\right) =-\mathbf{J}\left( -%
\mathbf{r},t\right) ,\ \ \mathcal{I~}\rho \left( \mathbf{r},t\right) =\rho
\left( -\mathbf{r},t\right) \ \ \ \ ~~
\end{align}

To see the effect of $\mathcal{I}$ on momentum-frequency domain quantities,
taking the Fourier transform of the parity-reversed field $\mathbf{E}\left( -%
\mathbf{r},t\right) $ leads to
\begin{align}
\mathcal{I\ }\mathbf{E}\left( \mathbf{k},\omega \right) & \equiv
\int\nolimits_{\mathbf{r}}\int\nolimits_{t}\mathcal{I~}\mathbf{E}\left(
\mathbf{r},t\right) e^{i\mathbf{k}\cdot \mathbf{r}}e^{-i\omega t}d\mathbf{r}%
dt \\
& =-\int\nolimits_{\mathbf{r}}\int\nolimits_{t}\mathbf{E}\left( -\mathbf{r}%
,t\right) e^{i\mathbf{k}\cdot \mathbf{r}}e^{-i\omega t}d\mathbf{r}dt
\nonumber \\
& =-\int\nolimits_{\mathbf{x}}\int\nolimits_{t}\mathbf{E}\left( \mathbf{x}%
,\tau \right) e^{-i\mathbf{k}\cdot \mathbf{x}}e^{-i\omega t}d\mathbf{x}dt
\nonumber \\
& =-\mathbf{E}\left( -\mathbf{k},\omega \right) =-\mathbf{E}^{\ast }\left(
\mathbf{k},-\omega \right)   \nonumber
\end{align}
Therefore, the prescription is that applying $\mathcal{I}$ in the space
domain is equivalent to taking complex conjugate and reversing frequency
(or, not taking complex conjugate and reversing momentum$\,$). For the
magnetic field, we get $\mathcal{I~}\mathbf{H}\left( \mathbf{k},\omega
\right) =\mathbf{H}^{\ast }\left( \mathbf{k},-\omega \right) $. For the source-free case,
\begin{align}
\mathcal{I\ }\mathbf{E}\left( \mathbf{k},\omega \right) & \equiv
\int\nolimits_{\mathbf{r}}\int\nolimits_{t=-\infty }^{t=\infty }\mathcal{I~%
}\mathbf{E}\left( \mathbf{r},t\right) e^{i\mathbf{k}\left( \omega \right)
\cdot \mathbf{r}}e^{-i\omega t}d\mathbf{r}dt \\
& =-\int\nolimits_{\mathbf{r}}\int\nolimits_{t}\mathbf{E}\left( -\mathbf{r}%
,t\right) e^{i\mathbf{k}\left( \omega \right) \cdot \mathbf{r}}e^{-i\omega
t}d\mathbf{r}dt  \nonumber \\
& =\int\nolimits_{-\mathbf{x}}\int\nolimits_{t}\mathbf{E}\left( \mathbf{x}%
,t\right) e^{-i\mathbf{k}\left( \omega \right) \cdot \mathbf{x}}e^{-i\omega
t}d\mathbf{x}dt  \nonumber \\
& =-\mathbf{E}\left( -\mathbf{k}\left( \omega \right) \right)   \nonumber \\
& =-\binom{\mathbf{E}^{\ast }\left( \mathbf{k}\left( -\omega \right) \right)
\text{ \ \ if }\mathbf{k}\left( \omega \right) =\mathbf{k}^{\text{even}}\left( \omega
\right) }{\mathbf{E}^{\ast }\left( -\mathbf{k}\left( -\omega \right) \right)
\text{ \ \ if }\mathbf{k}\left( \omega \right) =\mathbf{k}^{\text{odd}}\left( \omega
\right)  }.  \nonumber
\end{align}
Therefore,
\begin{align}
\mathcal{I\ }\boldsymbol{f}_{n,\mathbf{k}\left( \omega \right) }& =\left(
\begin{array}{c}
-\mathbf{E}^{\ast }\left( \mathbf{k}\left( -\omega \right) \right)  \\
\mathbf{H}^{\ast }\left( \mathbf{k}\left( -\omega \right) \right)
\end{array}%
\right) =-\,\,\boldsymbol{\mathrm{T}}_{6\times 6}\cdot \boldsymbol{f}_{n,-%
\mathbf{k}\left( \omega \right) }  \notag \\
& =\,-\,\binom{\boldsymbol{\mathrm{T}}_{6\times 6}\cdot \boldsymbol{f}_{n,%
\mathbf{k}\left( -\omega \right) }^{\ast }\text{ \ if }\mathbf{k}\left(
\omega \right) =\mathbf{k}^{\text{even}}\left( \omega \right) }{\boldsymbol{%
\mathrm{T}}_{6\times 6}\cdot \boldsymbol{f}_{n,-\mathbf{k}\left( -\omega
\right) }\text{ \ if }\mathbf{k}\left( \omega \right) =\mathbf{k}^{\text{odd}%
}\left( \omega \right) \text{.}}  \label{IEF}
\end{align}
If the system is $\mathcal{I}$-invariant, $-\boldsymbol{\mathrm{T}}_{6\times
6}\cdot \boldsymbol{f}_{n,-\mathbf{k}\left( \omega \right) }=$ $e^{i\zeta
\left( \mathbf{k}\right) }\boldsymbol{f}_{n,\mathbf{k}\left( \omega \right) }
$.

Starting with (\ref{KD}) the effect of $\mathcal{I}$ can be seen by considering that in the space
domain $\mathbf{w}$ is a convolution of the inverse spatial transforms of
the two terms. Space inversion of each term leads to, by the convolution theorem, the
product of the individual transforms evaluated at $-\mathbf{k}$. Using (\ref{SI}),
\begin{align}
\mathbf{w}_{n,\mathbf{k}\left( \omega \right) }=& \boldsymbol{\mathrm{T}}%
_{6\times 6}\cdot \left[ \partial _{\omega }\left( \omega \boldsymbol{%
\mathrm{M}}\left( \omega ,-\mathbf{k}\right) \right) \right] ^{1/2}\cdot
\boldsymbol{\mathrm{T}}_{6\times 6}  \notag \\
& \cdot \left( \,\boldsymbol{\mathrm{T}}_{6\times 6}\cdot \boldsymbol{f}_{n,-%
\mathbf{k}\left( \omega \right) }\right)   \notag \\
=& \boldsymbol{\mathrm{T}}_{6\times 6}\cdot \mathbf{w}_{n,-\mathbf{k}\left(
\omega \right) }
\end{align}
The effect of inversion on the Berry connection is
\begin{align}
\mathcal{I\ }\boldsymbol{\mathrm{A}}_{n}(\mathbf{k})& =i\left( \mathcal{I~}%
\mathbf{w}_{n,\mathbf{k}}\right) ^{\dagger }\cdot \nabla _{\mathbf{k}}\left(
\mathcal{I~}\mathbf{w}_{n,\mathbf{k}}\right)  \\
& =i\,\,\left( \boldsymbol{\mathrm{T}}_{6\times 6}\cdot \mathbf{w}_{n,-%
\mathbf{k}}\right) ^{\dagger }\cdot \nabla _{\mathbf{k}}\left( \boldsymbol{%
\mathrm{T}}_{6\times 6}\cdot \mathbf{w}_{n,-\mathbf{k}}\right)  \\
& =-i\mathbf{w}_{n,-\mathbf{k}}^{\dag }\cdot \nabla _{-\mathbf{k}}\mathbf{w}%
_{n,-\mathbf{k}}=-\mathbf{A}_{n}\left( -\mathbf{k}\right)
\end{align}
and that, if a system is $\mathcal{I}$-invariant,
\begin{equation}
-\boldsymbol{\mathrm{A}}_{n}(-\mathbf{k})=\boldsymbol{\mathrm{A}}_{n}(%
\mathbf{k})+\nabla _{\mathbf{k}}\xi \left( \mathbf{k}\right) ,
\end{equation}%
From this, for the Berry curvature,
\begin{align}
\mathcal{I~}\boldsymbol{\mathrm{F}}_{n}(\mathbf{k})& =\nabla _{\mathbf{k}%
}\times \left( \mathcal{I~}\mathbf{A}_{n}\left( \mathbf{k}\right) \right)  \\
& =\nabla _{\mathbf{k}}\times \left( -\mathbf{A}_{n}\left( -\mathbf{k)}%
\right) \right)   \notag \\
& =-\nabla _{-\mathbf{k}}\times \left( -\mathbf{A}_{n}\left( -\mathbf{k)}%
\right) \right)   \notag \\
& =\boldsymbol{\mathrm{F}}_{n}(-\mathbf{k})  \notag
\end{align}
and so if a system is invariant under $\mathcal{I}$,
\[
\boldsymbol{\mathrm{F}}_{n}(\mathbf{k})=\boldsymbol{\mathrm{F}}_{n}(-\mathbf{%
k}).
\]%
For the Chern number, $\mathcal{IC=C}$. Obviously, if a system is
invariant
under both $\mathcal{T}$ and $\mathcal{I}$, $\boldsymbol{\mathrm{F}}_{n}(%
\mathbf{k})=\mathbf{0}$.

\subsubsection{$\mathcal{T}$ and $\mathcal{I}$ relations for material parameters, and reciprocity}

\paragraph{Time reversal}

From (\ref{TEF}), $\mathcal{T\ }\boldsymbol{f}_{n,\mathbf{k}\left( \omega
\right) }=\boldsymbol{f}_{n}^{\text{TR}}=\,\,\boldsymbol{\mathrm{T}}%
_{6\times 6}\cdot \boldsymbol{f}_{n,-\mathbf{k}\left( \omega \right) }^{\ast
}$, and so for $\boldsymbol{g}_{n,\mathbf{k}\left( \omega \right) }=[%
\boldsymbol{\mathrm{D}}~~\boldsymbol{\mathrm{B}}]^{\mathrm{T}}$ we have $%
\boldsymbol{g}_{n}^{\text{\textrm{TR}}}=\boldsymbol{\mathrm{T}}_{6\times
6}\cdot \boldsymbol{g}_{n,-\mathbf{k}\left( \omega \right) }^{\ast }$.
Considering $\boldsymbol{g}_{n,\mathbf{k}\left( \omega \right) }=\boldsymbol{%
\mathrm{M}}\left( \omega ,\mathbf{k}\right) \cdot \boldsymbol{f}_{n,\mathbf{k%
}\left( \omega \right) }$, conjugating and changing the sign of momentum,
and multiplying by the Poynting vector reversing operator yields
\begin{align}
&\boldsymbol{\mathrm{T}}_{6\times 6}\cdot \boldsymbol{g}_{n,-\mathbf{k}\left(
\omega \right) }^{\ast } =\boldsymbol{\mathrm{T}}_{6\times 6}\cdot
\boldsymbol{\mathrm{M}}\left( \omega ,-\mathbf{k}\right) ^{\ast }\cdot
\boldsymbol{f}_{n,-\mathbf{k}\left( \omega \right) }^{\ast } \\
& =\left\{ \boldsymbol{\mathrm{T}}_{6\times 6}\cdot \boldsymbol{\mathrm{M}}%
\left( \omega ,-\mathbf{k}\right) ^{\ast }\cdot \boldsymbol{\mathrm{T}}%
_{6\times 6}\right\} \cdot \boldsymbol{\mathrm{T}}_{6\times 6}\cdot
\boldsymbol{f}_{n,-\mathbf{k}\left( \omega \right) }^{\ast }  \notag
\end{align}%
such that
\begin{equation}
\boldsymbol{g}_{n}^{\mathrm{TR}}=\boldsymbol{\mathrm{M}}^{\mathrm{TR}}\cdot
\boldsymbol{f}_{n}^{\mathrm{TR}}
\end{equation}%
where
\begin{eqnarray}
\boldsymbol{\mathrm{M}}^{\mathrm{TR}}\left( \omega ,\mathbf{k}\right)  &=&%
\boldsymbol{\mathrm{T}}_{6\times 6}\cdot \boldsymbol{\mathrm{M}}^{\ast
}\left( \omega ,-\mathbf{k}\right) \cdot \boldsymbol{\mathrm{T}}_{6\times 6}
\label{TR1} \\
&=&\left(
\begin{array}{cc}
\varepsilon _{0}\boldsymbol{\varepsilon }^{\ast }\left( \omega ,-\mathbf{k}\right)
& -\frac{1}{c}\boldsymbol{\xi }^{\ast }\left( \omega ,-\mathbf{k}\right)  \\
-\frac{1}{c}\boldsymbol{\varsigma }^{\ast }\left( \omega ,-\mathbf{k}\right)
& \mu _{0}\boldsymbol{\mu }^{\ast }\left( \omega ,-\mathbf{k}\right)
\end{array}%
\right)   \notag
\end{eqnarray}
is the time reversed material constitutive tensor. If we have $\boldsymbol{%
\mathrm{M}}=\boldsymbol{\mathrm{M}}^{\mathrm{TR}}$ then the medium is
time-reversal symmetric/invariant. Note that a lossy medium is never $\mathcal{T}$-invariant.

\paragraph{Space Inversion}

For the inversion operator, from (\ref{IEF}) $\mathcal{I\ }\boldsymbol{f}_{n,%
\mathbf{k}\left( \omega \right) }=\boldsymbol{f}_{n}^{\text{\textrm{I}}%
}=\,-\,\boldsymbol{\mathrm{T}}_{6\times 6}\cdot \boldsymbol{f}_{n,-\mathbf{k}%
\left( \omega \right) }$, and so for $\boldsymbol{g}_{n,\mathbf{k}\left(
\omega \right) }=[\boldsymbol{\mathrm{D}}~~\boldsymbol{\mathrm{B}}]^{\mathrm{%
T}}$ we have $\boldsymbol{g}_{n}^{\text{\textrm{I}}}=-\boldsymbol{\mathrm{T}}%
_{6\times 6}\cdot \boldsymbol{g}_{n,-\mathbf{k}\left( \omega \right) }$.
From $\boldsymbol{g}_{n,\mathbf{k}\left( \omega \right) }=\boldsymbol{%
\mathrm{M}}\left( \omega ,\mathbf{k}\right) \cdot \boldsymbol{f}_{n,\mathbf{k%
}\left( \omega \right) }$, changing the sign of momentum and multiplying by $%
\left( -\,\boldsymbol{\mathrm{T}}_{6\times 6}\right) $ results in
\begin{align}
& -\boldsymbol{\mathrm{T}}_{6\times 6}\cdot \boldsymbol{g}_{n,-\mathbf{k}%
\left( \omega \right) }=-\boldsymbol{\mathrm{T}}_{6\times 6}\cdot
\boldsymbol{\mathrm{M}}\left( \omega ,-\mathbf{k}\right) \cdot \boldsymbol{f}%
_{n,-\mathbf{k}\left( \omega \right) } \\
& =\left\{ \boldsymbol{\mathrm{T}}_{6\times 6}\cdot \left( -\boldsymbol{%
\mathrm{M}}\left( \omega ,-\mathbf{k}\right) \right) \cdot \boldsymbol{%
\mathrm{T}}_{6\times 6}\right\} \cdot \boldsymbol{\mathrm{T}}_{6\times
6}\cdot \boldsymbol{f}_{n,-\mathbf{k}\left( \omega \right) }  \notag
\end{align}
such that
\begin{equation}
\boldsymbol{g}_{n}^{\mathrm{I}}=\boldsymbol{\mathrm{M}}^{\mathrm{I}}\cdot
\boldsymbol{f}_{n}^{\mathrm{I}}
\end{equation}%
where
\begin{eqnarray}
\boldsymbol{\mathrm{M}}^{\mathrm{I}}\left( \omega \right)  &=&\boldsymbol{%
\mathrm{T}}_{6\times 6}\cdot \left( -\boldsymbol{\mathrm{M}}\left( \omega ,-%
\mathbf{k}\right) \right) \cdot \boldsymbol{\mathrm{T}}_{6\times 6} \label{SI} \\
&=&\left(
\begin{array}{cc}
\varepsilon _{0}\boldsymbol{\varepsilon }\left( \omega ,-\mathbf{k}\right)  & -%
\frac{1}{c}\boldsymbol{\xi }\left( \omega ,-\mathbf{k}\right)  \\
-\frac{1}{c}\boldsymbol{\varsigma }\left( \omega ,-\mathbf{k}\right)  & \mu
_{0}\boldsymbol{\mu }\left( \omega ,-\mathbf{k}\right)
\end{array}%
\right)   \notag
\end{eqnarray}
is the space inverted material constitutive tensor. Chirality ($\boldsymbol{%
\xi },\boldsymbol{\xi }\neq 0$), for example, breaks inversion symmetry.

\paragraph{Reciprocity}

The Maxwell's equations in the frequency domain in the presence of
an excitation may be written in a compact form as:
\begin{eqnarray}
\hat N \cdot {\bf{F}} = \omega {\bf{M}} \cdot {\bf{F}} + i{\bf{j}}
\end{eqnarray}
where $\hat N$ is a differential operator and $\bf{M}$ is the
material matrix operator \cite{Mario2}. For spatially dispersive
media the action of the material matrix on the electromagnetic
fields ${\bf{F}} = {\left( {\begin{array}{*{20}{c}}
{\bf{E}}&{\bf{H}}
\end{array}} \right)^T}$ should be understood as a spatial
convolution. The six-vector ${\bf{j}} = {\left(
{\begin{array}{*{20}{c}} {{{\bf{j}}_e}}&{{{\bf{j}}_m}}
\end{array}} \right)^T}$ is written in terms of the electric current
density ${{\bf{j}}_e}$ and of the magnetic current density
${{\bf{j}}_m}$ (for the sake of generality we consider that both
excitations are possible). In conventional media, the reciprocity
theorem establishes that if $\bf{F'}$ and $\bf{{F''}}$ are the
fields radiated by the localized (in space) currents $\bf{j'}$ and
$\bf{{j''}}$, respectively, then:
\begin{eqnarray}
\int {{\bf{j'}} \cdot {{\bf{T}}_{6 \times 6}} \cdot {\bf{F''}}}  -
{\bf{j''}} \cdot {{\bf{T}}_{6 \times 6}} \cdot {\bf{F'}}dV = 0.
\label{ApRec}
\end{eqnarray}
In the following, we study in which conditions this result
generalizes to spatially dispersive media. As a starting point, we
note that Parseval's theorem establishes that the condition
\eqref{ApRec} is equivalent to $\int {{{{\bf{j'}}}_{\bf{k}}} \cdot
{{\bf{T}}_{6 \times 6}} \cdot {{{\bf{F''}}}_{ {\bf{-k}}}}}  -
{{\bf{j''}}_{ {\bf{-k}}}} \cdot {{\bf{T}}_{6 \times 6}} \cdot
{{\bf{F'}}_{\bf{k}}}{d^3}{\bf{k}} = 0$, where ${{\bf{F'}}_{\bf{k}}}$
represents the Fourier transform of ${\bf{F'}}$, etc. The spectral
domain fields satisfy $\hat N\left( {\bf{k}} \right) \cdot
{{\bf{F}}_{\bf{k}}} = \omega {\bf{M}}\left( {\omega ,{\bf{k}}}
\right) \cdot {{\bf{F}}_{\bf{k}}} + i{{\bf{j}}_{\bf{k}}}$ which
generalizes the formulation of Sec. \ref{SecMax}. Hence, it follows
that:
\begin{eqnarray}
\begin{array}{l}
\int {{d^3}{\bf{k}}\left\{ {{{{\bf{F''}}}_{ - {\bf{k}}}} \cdot {{\bf{T}}_{6 \times 6}} \cdot \left[ {\hat N\left( {\bf{k}} \right) - {\bf{M}}\left( {\omega ,{\bf{k}}} \right)} \right] \cdot {{{\bf{F'}}}_{\bf{k}}}} \right.} \\
{\rm{        }}\left. { - {{{\bf{F'}}}_{\bf{k}}} \cdot {{\bf{T}}_{6
\times 6}} \cdot \left[ {\hat N\left( { {\bf{-k}}} \right) -
{\bf{M}}\left( {\omega , - {\bf{k}}} \right)} \right] \cdot
{{{\bf{F''}}}_{ {\bf{-k}}}}} \right\} = 0
\end{array}
\end{eqnarray}
Since the sources are arbitrary the above equation can be satisfied
only when:
\begin{eqnarray}
\begin{array}{l}
{{\bf{T}}_{6 \times 6}} \cdot \left[ {\hat N\left( {\bf{k}} \right) - {\bf{M}}\left( {\omega ,{\bf{k}}} \right)} \right] = \\
{\rm{       }}{\left[ {{{\bf{T}}_{6 \times 6}} \cdot \left[ {\hat
N\left( { - {\bf{k}}} \right) - {\bf{M}}\left( {\omega , - {\bf{k}}}
\right)} \right]} \right]^T}
\end{array}
\end{eqnarray}

Using the explicit expression of the matrix $\hat N\left( {\bf{k}}
\right)$ (Eq. \eqref{matMN}) it is found after some manipulations
that a material is reciprocal only when the respective material
matrix satisfies:
\begin{eqnarray}
{\bf{M}}\left( {\omega ,{\bf{k}}} \right) = {{\bf{T}}_{6 \times 6}}
\cdot {{\bf{M}}^T}\left( {\omega , - {\bf{k}}} \right) \cdot
{{\bf{T}}_{6 \times 6}}.
\end{eqnarray}
Comparing the requirements for TR symmetry and reciprocity, we
conclude that a reciprocal material is $\mathcal{T}$-invariant, and
a $\mathcal{T}$-invariant material is reciprocal, provided that
\begin{equation}
\boldsymbol{\mathrm{M}}^{\ast }=\boldsymbol{\mathrm{M}}^{\mathrm{T}%
}\rightarrow
\boldsymbol{\mathrm{M}}=\boldsymbol{\mathrm{M}}^{\dagger },
\end{equation}%
i.e., that the material tensor is Hermitian. Since all lossless
materials must have Hermitian matrix representations, in the
loss-free case a reciprocal medium also has TR symmetry, and vice
versa.



\begin{thebibliography}{1}
\bibitem{griffithsA} One widely used example is moving a swinging pendulum from one point on a sphere, along some contour, back to the original point - the geometric phase is given by the solid angle subtended by the path of movement. See \cite{griffiths}.
\end{thebibliography}
\end{document}